\documentclass[lettersize,journal]{IEEEtran}
\usepackage{cite}
\usepackage{amsfonts,amsmath,color,amssymb,amsxtra,amsbsy,floatflt}
\usepackage{algorithmic}
\usepackage{graphicx}
\usepackage{textcomp}
\usepackage{xcolor}
\def\BibTeX{{\rm B\kern-.05em{\sc i\kern-.025em b}\kern-.08em
    T\kern-.1667em\lower.7ex\hbox{E}\kern-.125emX}}

\usepackage{enumitem}
\newlist{myenum}{enumerate}{3}
\setlist[myenum,1]{label*=\arabic*)}
\setlist[myenum,2]{label=\arabic{myenumi}.\arabic*)}
\setlist[myenum,3]{label=\arabic{myenumi}.\arabic{myenumii}.\arabic*)}

\def\1{\mathbf{1}}

\usepackage[acronym]{glossaries}
\usepackage{subfigure}
\usepackage[ruled,vlined,linesnumbered]{algorithm2e}
\usepackage{placeins}
\usepackage{epstopdf}
\usepackage{enumitem}

\usepackage{tikz}
\usetikzlibrary{automata, arrows, positioning,calc}

\usepackage{multirow}
\usepackage[export]{adjustbox}
\usepackage{xfrac}
\usepackage[hidelinks]{hyperref}

\newcommand{\smallminus}{\scalebox{0.5}[1.0]{$-$}}

\newcommand{\name}{{DARIO}\xspace}

\usepackage{fancyhdr}
\usepackage{tcolorbox}
\usepackage{tabularx}
\usepackage{booktabs}

\begin{document}

\thispagestyle{empty} 

\vfill 

\begin{center}
\begin{tcolorbox}[
    colback=gray!10, colframe=black, fonttitle=\bfseries,
    width=0.9\textwidth, boxrule=1pt, arc=5pt, outer arc=5pt,
    boxsep=10pt, left=10pt, right=10pt, top=10pt, bottom=10pt
]
\centering
\textbf{THIS IS AN AUTHOR-CREATED POSTPRINT VERSION.}

\vspace{0.3cm}

\textbf{Disclaimer:}  
This work has been accepted for publication in \textit{IEEE Transactions on Wireless Communications}.  

\vspace{0.3cm}

\textbf{Copyright:}  
© 2026 IEEE. Personal use of this material is permitted. Permission from IEEE must be obtained for all other  
uses, in any current or future media, including reprinting/republishing this material for advertising or  
promotional purposes, creating new collective works, for resale or redistribution to servers or lists, or reuse of  
any copyrighted component of this work in other works.

\vspace{0.3cm}

\textbf{DOI:} \href{https://doi.org/10.1109/TWC.2026.3672338}{10.1109/TWC.2026.3672338}

\end{tcolorbox}
\end{center}

\vfill 

\clearpage
\setcounter{page}{1}

\newacronym{3GPP}{3GPP}{3rd Generation Partnership Project}

\newacronym{5G}{5G}{5th Generation}
\newacronym{5G-ACIA}{5G-ACIA}{5G Alliance for Connected Industries and Automation}
\newacronym{5G-NR}{5G-NR}{5G New Radio}
\newacronym{6G}{6G}{Sixth Generation}

\newacronym{AMC}{AMC}{Adaptive Modulation and Coding}
\newacronym{AC}{AC}{Admission Control}
\newacronym{AGV}{AGV}{Automated Guided Vehicle}
\newacronym{AR}{AR}{Augmented Reality}

\newacronym{BLER}{BLER}{Block Error Rate}
\newacronym{BWP}{BWP}{Bandwidth Part}
\newacronym{BS}{BS}{Base Station}
\newacronym{BSS}{BSS}{Business Support System}

\newacronym{CDF}{CDF}{Cumulative Distribution Function}
\newacronym{CCDF}{CCDF}{Complementary Cumulative Distribution Function}
\newacronym{CDMA}{CDMA}{Code Division Multiple Access}
\newacronym{CTMC}{CTMC}{Continuos-Time Markov Chain}
\newacronym{CSI}{CSI}{Channel State Information}
\newacronym{CP}{CP}{Control Plane}
\newacronym{CQI}{CQI}{Channel Quality Indicator}
\newacronym{CU}{CU}{Centralized Unit}

\newacronym{DL}{DL}{downlink}
\newacronym{DNC}{DNC}{Deterministic Network Calculus}
\newacronym{DRP}{DRP}{Dynamic Resource Provisioning}
\newacronym{DRL}{DRL}{Deep Reinforcement Learnning}
\newacronym{DU}{DU}{Distributed Unit}

\newacronym{eMBB}{eMBB}{enhanced Mobile Broadband}
\newacronym{ETSI}{ETSI}{European Telecommunication Standards Institute}
\newacronym{EBB}{EBB}{Exponentially Bounded Burstiness}
\newacronym{EBF}{EBF}{Exponentially Bounded Fluctuation}
\newacronym{E2E}{E2E}{End-to-End}
\newacronym{EDF}{EDF}{Earliest Deadline First}
\newacronym{EM}{EM}{Expectation-Maximization}

\newacronym{FCFS}{FCFS}{First-come First-served}
\newacronym{FIFO}{FIFO}{First In First Out}

\newacronym{GBR}{GBR}{Guaranteed Bit Rate}
\newacronym{GMM}{GMM}{Gaussian Mixture Model}
\newacronym{GSMA}{GSMA}{Global System for Mobile Communications Association}
\newacronym{GST}{GST}{Generic Network Slice Template}
\newacronym{gNB}{gNB}{Next generation NodeB}

\newacronym{HDR}{HDR}{High Data Rate}

\newacronym{ITU}{ITU}{International Telecommunication Union}
\newacronym{IoT}{IoT}{Internet of Things}
\newacronym{ILP}{ILP}{Integer Linear Programming}
\newacronym{ICIC}{ICIC}{Inter-Cell Interference Cancellation}
\newacronym{ISAC}{ISAC}{Integrated Sensing and Communication}

\newacronym{LA}{LA}{Link Adaptation}
\newacronym{LOS}{LoS}{Line-of-Sight}
\newacronym{LSTM}{LSTM}{Long Short-Term Memory}
\newacronym{LTE}{LTE}{Long Term Evolution}

\newacronym{MAC}{MAC}{Medium Access Control}
\newacronym{MEC}{MEC}{Multi-access Edge Computing}
\newacronym{MCS}{MCS}{Modulation and Coding Scheme}
\newacronym{MDN}{MDN}{Mixture Density Network}
\newacronym{MGF}{MGF}{Moment Generating Function}
\newacronym{MIMO}{MIMO}{Multiple Input Multiple Output}
\newacronym{MISO}{MISO}{Multiple Input Single Output}
\newacronym{MRC}{MRC}{Maximal Ratio Combining}
\newacronym{ML}{ML}{Machine Learning}
\newacronym{MNO}{MNO}{Mobile Network Operator}
\newacronym{mMTC}{mMTC}{Machine Type Communication}
\newacronym{MSE}{MSE}{Mean Squared Error}
\newacronym{mURLLC}{mURLLC}{massive ultra-Reliable Low Latency Communication}

\newacronym{NE}{NE}{Nash Equilibrium}
\newacronym{NEST}{NEST}{Network Slice Type}
\newacronym{NIP}{NIP}{Non-linear Integer Programming}
\newacronym{NFMF}{NFMF}{Network Function Management Function}
\newacronym{NFV}{NFV}{Network Function Virtualization}
\newacronym{NG-RAN}{NG-RAN}{Next Generation - RAN}
\newacronym{NLOS}{NLoS}{Non-Line-of-Sight}
\newacronym{NN}{NN}{Neural Network}
\newacronym{NSO}{NSO}{Network Slice Orchestrator}
\newacronym{NSMF}{NSMF}{Network Slice Management Function}
\newacronym{NSSMF}{NSSMF}{Network Slice Subnet Management Function}
\newacronym{NR}{NR}{New Radio}

\newacronym{OFDMA}{OFDMA}{Orthogonal Frequency-Division Multiple Access}
\newacronym{O-RAN}{O-RAN}{Open Radio Access Network}
\newacronym{PDF}{PDF}{Probability Density Function}
\newacronym{PMF}{PMF}{Probability Mass Function}
\newacronym{PRB}{PRB}{Physical Resource Block}
\newacronym{P-NEST}{P-NEST}{private NEST}

\newacronym{QoS}{QoS}{Quality of Service}

\newacronym{RAN}{RAN}{Radio Access Network}
\newacronym{RB}{RB}{Resource Block}
\newacronym{RBG}{RBG}{Resource Block Group}
\newacronym{RIC}{RIC}{RAN Intelligent Controller}
\newacronym{RIS}{RIS}{Reconfigurable Intelligent Surfaces}
\newacronym{RRM}{RRM}{Radio Resource Management}
\newacronym{RSMA}{RSMA}{Rate Splitting Multiple Access}
\newacronym{RSRP}{RSRP}{Received Signal Received Power}
\newacronym{RSRQ}{RSRQ}{Received Signal Received Quality}
\newacronym{RSSI}{RSSI}{Received Signal Strength Indication}
\newacronym{RT}{RT}{Real Time}
\newacronym{RU}{RU}{Radio Unit}

\newacronym{SDO}{SDO}{Standards Developing Organization}
\newacronym{SINR}{SINR}{Signal-to-Interference-plus-Noise Ratio}
\newacronym{SLA}{SLA}{Service Level Agreement}
\newacronym{SNC}{SNC}{Stochastic Network Calculus}
\newacronym{SNR}{SNR}{Signal-to-Noise Ratio}
\newacronym{S-NEST}{S-NEST}{standardized NEST}
\newacronym{SIMO}{SIMO}{Single-Input Multiple-Output}
\newacronym{SMO}{SMO}{Service Management and Orchestration}

\newacronym{TTI}{TTI}{Transmission Time Interval}

\newacronym{UE}{UE}{User Equipment}
\newacronym{UL}{UL}{Uplink}
\newacronym{UP}{UP}{User Plane}
\newacronym{uRLLC}{uRLLC}{ultra-Reliable Low Latency Communication}

\newacronym{V2X}{V2X}{Vehicle-to-Everything}
\newacronym{VBR}{VBR}{Variable Bit Rate}
\newacronym{VR}{VR}{Virtual Reality}
\newacronym{vRAN}{vRAN}{virtualized RAN}
\newacronym{vBS}{vBS}{virtualized Base Station}
\newacronym{VS}{VS}{Validation Scenario}

\newacronym{WiMAX}{WiMAX}{Worldwide Interoperability for Microwave Access}
\newacronym{WCDMA}{WCDMA}{Wideband \gls{CDMA}}
\newacronym{WRR}{WRR}{Weighted Round Robin}

\title{RIS Control through the Lens of Stochastic Network Calculus: An O-RAN Framework for Delay-Sensitive 6G Applications

\thanks{Funding for open access charge: Universidad de Granada / CBUA. This work is part of the project PID2022-137329OB-C43 funded by MICIU/AEI/10.13039/501100011033 and by FEDER, EU, and also part of the project C-ING-306-UGR23 funded by Consejería de Universidad, Investigación e Innovación and by ERDF Andalusia Program 2021-2027. Furthermore, it has been partially supported by the MultiX project from Smart Networks and Services Joint Undertaking (SNS JU) under the European Union’s Horizon Europe research and innovation programme (Grant 101192521).

Oscar Adamuz-Hinojosa is with the Department of Signal Theory, Telematics and Communications, University of Granada, Granada, Spain (e-mail: \texttt{oadamuz@ugr.es}). Lanfranco Zanzi, Vincenzo Sciancalepore, and Xavier Costa-Perez are with NEC Laboratories Europe, Heidelberg, Germany. (e-mail: \texttt{\{name.surname\}@neclab.eu}). Xavier Costa-Perez is also with i2CAT Foundation and ICREA, Barcelona, Spain (e-mail: \texttt{xavier.costa@i2cat.net}). M. Di Renzo is with Universit\'e Paris-Saclay, CNRS, CentraleSup\'elec, Laboratoire des Signaux et Syst\`emes, 3 Rue Joliot-Curie, 91192 Gif-sur-Yvette, France. (marco.di-renzo@universite-paris-saclay.fr), and with King's College London, Department of Engineering - Centre for Telecommunications Research, WC2R 2LS London, United Kingdom (marco.di\_renzo@kcl.ac.uk).}

}

\author{ 
{Oscar Adamuz-Hinojosa},
{Lanfranco Zanzi},~\IEEEmembership{Member,~IEEE,}
{Vincenzo~Sciancalepore},~\IEEEmembership{Senior Member,~IEEE,}
{Marco Di Renzo},~\IEEEmembership{Fellow,~IEEE,}
{Xavier~Costa-Pérez},~\IEEEmembership{Senior Member,~IEEE}

}



\maketitle

\begin{abstract}
Reconfigurable Intelligent Surfaces (RIS) enable dynamic electromagnetic control for 6G networks, but existing control schemes lack responsiveness to fast-varying network conditions, limiting their applicability for ultra-reliable low latency communications. This work address uplink delay minimization in multi-RIS scenarios with heterogeneous per-user latency and reliability demands. We propose Delay-Aware RIS Orchestrator (\name{}), an O-RAN-compliant framework that dynamically assigns RIS devices to users within short time windows, adapting to traffic fluctuations to meet per-user delay and reliability targets. \name{} relies on a novel Stochastic Network Calculus (SNC) model to analytically estimate the delay bound for each possible user–RIS assignment under specific traffic and service dynamics. These estimations are used  by DARIO to formulate a Nonlinear Integer Program (NIP), for which an online heuristic provides near-optimal performance with low computational overhead. Extensive evaluations with simulations and real traffic traces show consistent delay reductions up to $95.7\%$ under high load or RIS availability.
\end{abstract}


\maketitle

\section{Introduction}
\label{sec:intro}

\gls{RIS} are key enablers for \gls{6G} networks, mitigating signal blockages and multipath fading by electronically reconfiguring the propagation environment~\cite{Saad2020}. By adjusting the phase and amplitude of reflected signals, \gls{RIS} devices establish \textit{controlled \gls{LOS} links} between \glspl{UE} and \glspl{BS}~\cite{Kishk2021}. This reconfigurability enables new forms of adaptive radio resource scheduling, particularly suited to delay-sensitive and mission-critical applications.

\textbf{Motivation.} In addition to advanced radio resource scheduling techniques~\cite{Anand2020}, the application of \gls{RIS} devices in delay-sensitive scenarios can serve as a complementary approach to enhance performance. In such scenarios, it is crucial to ensure the packet delay budget of each \gls{UE} is respected, with violation probabilities remaining below predefined thresholds~\cite{Fidler2010}. A possible approach to address this requirement is to dynamically adapt the \gls{RIS} configuration across time slots to serve different \glspl{UE}, each experiencing distinct channel conditions, traffic patterns, and delay requirements. This would require adaptive \gls{UE}–\gls{RIS} association mechanisms capable of tracking these fast and diverse environmental changes.

To support such adaptive mechanisms, \gls{SNC} offers a rigorous mathematical framework for modeling the uncertainties arising from time-varying channels and stochastic traffic arrivals~\cite{Fidler1}. \gls{SNC} enables analytical estimation of probabilistic delay bounds~\cite{Adamuz-Hinojosa-SNC}, which can be used to assess whether a given combination of \gls{UE}–\gls{RIS} associations satisfies the required delay constraints. 


The envisioned \gls{RIS} orchestrator framework, supported by \gls{SNC}-based delay modeling, aligns with the principles of \gls{O-RAN}, which promotes disaggregated, programmable, and intelligent control of RAN functions~\cite{Polese2023}. \gls{O-RAN} introduces distributed control through the Non-\gls{RT} and Near-\gls{RT} \glspl{RIC}, separating long-term policy optimization from short-term control decisions~\cite{Adamuz-Hinojosa-MAREA}.  Within this architecture, 
a dedicated \gls{RIS} orchestrator can leverage real-time network state information accessed via the \glspl{RIC} to perform adaptive \gls{UE}-\gls{RIS} association. This enables dynamic \gls{RIS} reconfiguration in response to changing traffic patterns, mobility, and delay constraints. 


\textbf{Related Works}.
Prior research on \gls{RIS} mainly targets system capacity or energy efficiency optimization, often overlooking the stringent delay requirements of \gls{uRLLC} services. Du et al.~\cite{Du2021} analyze a multicast \gls{DL} scenario where a single \gls{RIS} assists a multi-antenna \gls{BS}, jointly optimizing transmit covariance and phase shifts to maximize information-theoretic capacity under perfect \gls{CSI}. This static, throughput-oriented formulation does not capture stochastic latency or reliability trade-offs. Similarly, Guo et al.~\cite{Guo2020} maximize weighted sum-rate in multiuser \gls{MISO} \gls{DL} systems using fractional programming and block coordinate descent, later extended to imperfect \gls{CSI}, but remain focused on deterministic spectral-efficiency gains without accounting for queueing dynamics or time-varying associations. Related works by Du et al.~\cite{Du2021-2} and Liu et al.~\cite{Liu2023} employ iterative optimization techniques to maximize energy efficiency or sum-rate, respectively, yet similarly rely on static rate or power metrics and do not address probabilistic delay guarantees or dynamic user–\gls{RIS} reassignment. Xue et al.~\cite{Xue2024} study user association in dense mmWave multi-connectivity scenarios via power optimization, while Peng et al.~\cite{Peng2024} propose a two-timescale \gls{RIS}-aided \gls{uRLLC} scheme under finite blocklength using statistical \gls{CSI} to reduce overhead. However, latency in this latter work is modeled deterministically through short-packet rate expressions, without per-user delay guarantees. In a related line of research, Yao et al.~\cite{Yao-1,Yao-2} investigate hybrid multi-\gls{RIS}-aided secure \gls{ISAC} systems, jointly optimizing transmit waveforms, receive filters, and discrete reflection coefficients to enhance sensing \gls{SINR} and physical-layer security under imperfect \gls{CSI}. While these works incorporate multi-user and multi-\gls{RIS} architectures with robust optimization techniques, their formulations remain snapshot-based and performance-driven (e.g., \gls{SINR} maximization), without modeling stochastic traffic arrivals, queue dynamics, or probabilistic delay guarantees.

A smaller body of work considers delay-aware formulations. Mukherjee et al.~\cite{Mukherjee2022} and Xia et al.~\cite{Xia2022} investigate \gls{MEC}-enabled networks with a single \gls{RIS}, aiming to minimize maximum delay by jointly accounting for transmission and computation latency. These problems are solved per snapshot through convex relaxations, but do not model stochastic traffic arrivals or per-user reliability. In the radio interface domain, Almekhlafi et al.~\cite{Almekhlafi2022} and De Souza et al.~\cite{Souza2024} address coexistence of \gls{eMBB} and \gls{uRLLC} traffic, focusing on power allocation, precomputed \gls{RIS} phase sets, and frame-based scheduling. Reliability is assessed via outage probability rather than probabilistic delay bounds, and all assume a static single-\gls{RIS} setup. Liu et al.~\cite{Liu2021_uplinkMultiplexing} study \gls{UL} multiplexing for joint \gls{eMBB}/\gls{uRLLC} services with \gls{RIS}-assisted scheduling, but do not consider pure \gls{uRLLC} scenarios under multiple \gls{RIS} devices, nor delay evolution over time.

More recent studies extend \gls{RIS} techniques to increasingly dynamic settings with multiple \glspl{UE}, adaptive configurations, and emerging \gls{O-RAN} architectures. Soleymani et al.~\cite{WCL2024_RateRegion} analyze the rate region of \gls{uRLLC} broadcast channels under different \gls{RIS} architectures, comparing diagonal and beyond-diagonal designs under reliability constraints. Their follow-up work~\cite{Soleymani2024_RSMA} incorporates \gls{RSMA} into \gls{MIMO} \gls{RIS}-aided systems with finite blocklength coding to improve rate–reliability trade-offs. Both formulations evaluate achievable rate and packet error probability under static user–\gls{RIS} topologies and perfect \gls{CSI}, without modeling queue dynamics or probabilistic delay bounds.

Beyond differences in optimization objectives, existing \gls{RIS} frameworks are not directly applicable to probabilistic delay-bound optimization without altering their core assumptions. They are predominantly based on snapshot-driven designs with quasi-static channels, fixed user--\gls{RIS} associations, and, in some cases, deterministic service rates, which prevents the modeling of queue dynamics, stochastic traffic arrivals, and delay accumulation over time. As a result, latency is typically treated either as a secondary metric via short-packet or outage-based formulations, or as a deterministic constraint tied to static topologies. Consequently, the resulting configurations (e.g., phase-shift or power settings) are computed offline or per frame and cannot be mapped to per-slot control policies with per-\gls{UE} probabilistic delay guarantees. Extending these approaches would require introducing explicit traffic and queueing models together with time-coupled decision-making, leading to fundamentally different problem formulations rather than direct adaptations. Moreover, none of the existing works jointly address multiple \gls{RIS} devices, user mobility, and time-varying traffic while enabling real-time adaptability under standardized \gls{O-RAN} architectures and an \gls{SNC}-based framework to provide per-\gls{UE} probabilistic delay guarantees.

{\it \bf Contributions.} 
This work is the first to design, implement, and evaluate a multi-\gls{RIS} network that serves mobile \glspl{UE} with heterogeneous delay budgets and violation probabilities. The main contributions are:

\textbf{C1)} Proposal of Delay-Aware RIS Orchestrator (\name{}), a \gls{O-RAN}-compliant \gls{RIS} orchestrator that dynamically configures \gls{RIS} devices across time slots to provide instantaneous \gls{LOS} conditions, improving \gls{UE} channels and reducing delay violation probabilities.

\textbf{C2)} Design a novel \gls{SNC}-based delay model capturing \gls{UL} rate variability due to dynamic \gls{UE}-\gls{RIS} configurations, allowing analytical estimation of delay violation probabilities under realistic traffic and channel dynamics.

\textbf{C3)} Formulation of the \gls{UE}–\gls{RIS} assignment as a \gls{NIP} problem, minimizing the ratio between experienced delay and delay bounds, aligned with real-time requirements.

\textbf{C4)} Design of a heuristic algorithm solving the \gls{NIP} with near-optimal performance and low computational complexity, suitable for real-time use in \gls{O-RAN} architecture.

\textbf{C5)} Comprehensive evaluation using simulations and real-world traffic traces, showing that \name{} consistently reduces delay across all scenarios, achieving up to $95.71\%$ improvement in high-load or high-\gls{RIS}-availability deployments.

The remainder of this paper is organized as follows. Section~\ref{sec:DARIOframework} describes \name's main functional blocks. Section~\ref{sec:system-model} defines the system model. Section~\ref{sec:SNCmodel} introduces the proposed \gls{SNC} model. Section~\ref{sec:ProblemFormulation} details the \gls{UE}-\gls{RIS} assignment and the proposed heuristics for solving it. Section~\ref{sec:perf_eval} evaluates \name’s performance, considering a scenario where \glspl{UE} generate traffic following a Poisson distribution. In contrast, Section~\ref{sec:real} assesses \name’s performance based on real traffic traces previously collected from a major mobile operator. Finally, Section~\ref{sec:Conclusions} summarizes the main conclusions.

\section{DARIO Framework}\label{sec:DARIOframework}
This section presents the proposed \name{} framework. First, it describes the framework architecture, its main functional entities, and the logical interfaces that enable interaction with existing \gls{O-RAN} control components. Then, the orchestration workflow of \name{} is detailed across the Non-\gls{RT} and Near-\gls{RT} control loops. Subsequently, the main implementation challenges for practical integration within \gls{O-RAN} are analyzed. Next, control-loop latency and signaling overhead are examined to assess the feasibility of the proposed orchestration within the \gls{O-RAN} temporal hierarchy. Finally, the scalability of \name{} is discussed, outlining how \name{} can be extended to multi-cell environments.

\subsection{Architecture, Functionalities, and Interfaces}\label{subsec:DARIO-architecture}

\begin{figure}[b!]
    \centering
    \includegraphics[width=0.88\columnwidth]{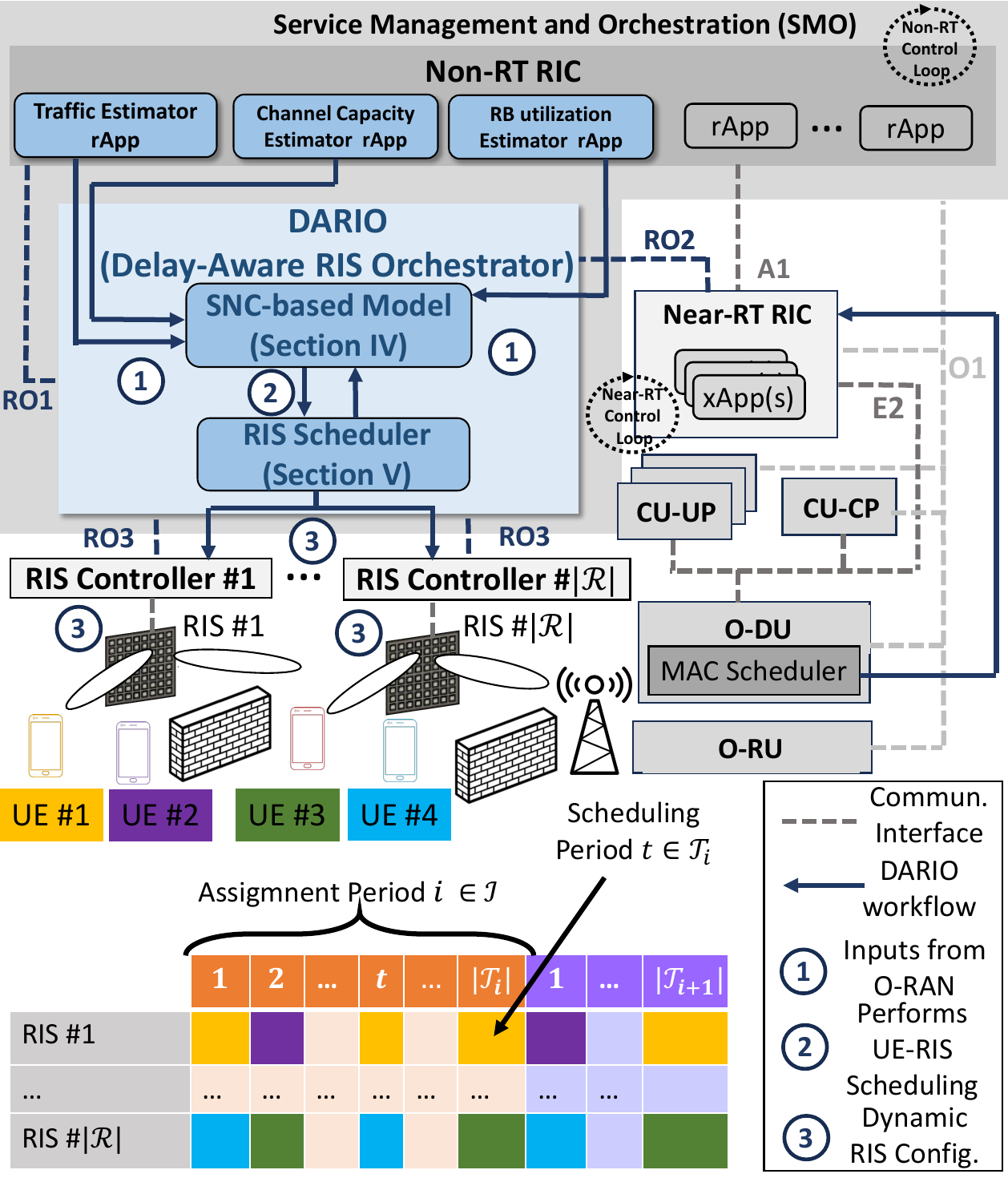}
    \caption{\name{} integrated within the \gls{O-RAN} architecture.}
    \label{fig:scenario}
\end{figure}

\name{}, illustrated in Fig.~\ref{fig:scenario},  is a delay and reliability aware orchestration framework designed to dynamically coordinate multiple \glspl{RIS} within the \gls{O-RAN} ecosystem. Its main function is to determine, for each \gls{UE} and at every scheduling slot, which single \gls{RIS} should be configured to provide this \gls{UE} the most favorable propagation conditions, together with the corresponding radio resource allocation. The selected \gls{RIS} configuration is dynamically updated across slots, allowing each \gls{UE} to be served by different \glspl{RIS} over time depending on the predicted channel and traffic states. The objective of \name{} is to minimize the per-packet transmission delay experienced by each \gls{UE} while ensuring the required reliability level across all scheduling intervals.

From an architectural perspective, \name{} operates as a separate orchestration module colocated with the \gls{SMO} and fully interoperable with existing \gls{O-RAN} components. It remains logically distinct from the Non-\gls{RT} \gls{RIC} to preserve compatibility with current specifications, while exploiting its analytics as input and providing optimized configuration policies as output to the Near-\gls{RT} control layer. This design enables seamless integration of \name{} into the \gls{O-RAN} management hierarchy, relying on the standard interfaces for coordination and introducing dedicated extensions to support \gls{RIS}-specific control. Internally, \name{} comprises two main entities:

\textbf{SNC Module:} Estimates the statistical delay bound $W$ for each feasible \gls{UE}--\gls{RIS} configuration and radio resource allocation using \gls{SNC} principles. The packet transmission delay $w$ in the radio interface is modeled as a random variable constrained by $\Pr[w > W] \leq \varepsilon$, where $\varepsilon$ is the maximum allowed violation probability.

\textbf{RIS Scheduler:} Iteratively interacts with the \gls{SNC} module to evaluate candidate \gls{UE}--\gls{RIS} mappings and RB allocations. It selects the configuration that minimizes the resulting delay bound $W$ while ensuring $W \leq W^{\text{th}}$ for each \gls{UE} and satisfying the reliability constraint $\Pr[w > W] \leq \varepsilon$.

To enable full integration with \gls{O-RAN} control entities, \name{} introduces three logical interfaces that extend those existing from \gls{O-RAN} specifications:

\textbf{\texttt{RO1}:} connects \name{} with the Non-\gls{RT} \gls{RIC}. It carries aggregated analytics produced by Non-\gls{RT} \emph{rApps}, including predicted traffic load per \gls{UE}, expected channel quality indicators, and radio resource utilization.

\textbf{\texttt{RO2}:} links \name{} with the Near-\gls{RT} \gls{RIC}. It conveys the optimization outcomes related to radio resource allocation and scheduling policies executed by \emph{xApps} in the Near-\gls{RT} domain.

\textbf{\texttt{RO3}:} interfaces \name{} with local \gls{RIS} controllers to deliver the specific configuration states computed by the \gls{RIS} scheduler, which are then translated by the \gls{RIS} controllers into element-level phase updates for each \gls{RIS}.

The analytics consumed through \texttt{RO1} originate from Non-\gls{RT} \emph{rApps} such as the \emph{Traffic Estimator}, \emph{Channel Capacity Estimator}, and \emph{RB Utilization} modules. These applications collect raw performance data from the network through the \texttt{O1} interface and produce aggregated indicators that serve as input for \name{} during each optimization cycle. The outputs generated by \name{} are transmitted through \texttt{RO2} to the Near-\gls{RT} \gls{RIC}, which manages radio resource allocation, and through \texttt{RO3} to local \gls{RIS} controllers, which apply the selected configuration states to their corresponding surfaces at the element level.

\subsection{Orchestration Workflow within O-RAN Control Loops}\label{subsec:DARIO-workflow}

The operation of \name{} follows the hierarchical control loops defined by the \gls{O-RAN} architecture, where optimization and actuation occur at different temporal scales. Two time parameters are defined: the assignment period $i \in \mathcal{I}$ of duration $I_{\scriptstyle time}$, typically a few seconds, and the scheduling period $t \in \mathcal{T}_i$ of duration $T_{\scriptstyle time}=I_{\scriptstyle time}/|\mathcal{T}_i|$, aligned with Near-\gls{RT} timescales between 10 and 1000~ms. During each assignment period $i \in \mathcal{I}$, \name{} executes its optimization process, while the corresponding \gls{UE}-\gls{RIS} configurations are enforced at every scheduling period $t \in \mathcal{T}_i$.

At the \textbf{Non-\gls{RT} control loop}, \name{} performs the end-to-end orchestration procedure: \textit{1)} Collects aggregated analytics from the Non-\gls{RT} \gls{RIC} through \texttt{RO1}, including predicted traffic demand per \gls{UE}, expected channel quality, and radio resource utilization generated by dedicated \emph{rApps}. \textit{2)} Executes an iterative process between the \gls{SNC} module and the \gls{RIS} scheduler to compute, for each scheduling period $t \in \mathcal{T}_i$, the optimal \gls{UE}–\gls{RIS} association that minimizes latency under reliability constraints, while defining a single RB allocation fixed throughout the current assignment period $i \in \mathcal{I}$. \textit{3)} Distributes the orchestration results to the corresponding control entities: the radio resource allocation policies are transmitted to the Near-\gls{RT} \gls{RIC} through \texttt{RO2}, and the \gls{RIS} configuration plan for each scheduling period is delivered directly to the local \gls{RIS} controllers through \texttt{RO3}.

At the \textbf{Near-\gls{RT} control loop}, the precomputed decisions are applied as follows: \textit{1)} The Near-\gls{RT} \gls{RIC} enforces the radio resource allocation and scheduling policies received from \name{} by means of its \emph{xApps}. \textit{2)} Each \gls{RIS} controller applies the corresponding phase-state settings to its \gls{RIS} devices via internal control links. These configurations are updated dynamically at the beginning of every scheduling period $t \in \mathcal{T}_i$ of duration $T_{\scriptstyle time}$. \textit{3)} The resulting \gls{RIS} phase profiles create controlled \gls{LOS} paths between the selected \glspl{UE} and serving \glspl{BS}, adapting over time to statistically maintain the latency and reliability levels defined by \name{}.

In this hierarchical process, \name{} performs optimization at the Non-\gls{RT} scale, while Near-\gls{RT} entities enforce RIS and scheduling configurations at millisecond granularity. Although PIN-diode \gls{RIS} elements can switch in sub-microseconds~\cite{Jiang2023}, the effective reconfiguration latency is dominated by channel estimation, computation, and O-RAN signaling~\cite{Polese2023}, making updates feasible only at Near-\gls{RT} intervals.

\subsection{Implementation Challenges for \gls{O-RAN} Integration}
\name{} is designed to operate within the \gls{O-RAN} framework, enabling hierarchical delay-aware \gls{RIS} orchestration across multiple time scales. However, real-world integration introduces practical challenges that must be addressed for reliable operation. Below, we outline key issues:

\textbf{Management Plane Latency}: Although \gls{RIS} devices support microsecond-scale switching~\cite{Jiang2023}, the end-to-end management pipeline, including Non-\gls{RT} and Near-\gls{RT} \glspl{RIC} as well as local \gls{RIS} Controllers, may introduce delays due to processing, queuing, and transmission. This challenge concerns ensuring that telemetry collection, Non-\gls{RT} optimization, and configuration dissemination comply with the temporal hierarchy defined by O-RAN. Our study demonstrates that \name{} meets the execution-time requirements of the Non-\gls{RT} control loop, but the design and dimensioning of the management network responsible for telemetry exchange and configuration delivery lie beyond the scope of this work. These mechanisms must be properly engineered in real implementations to sustain the required control-loop timing.

\textbf{Interface Compatibility}: Although \name{} leverages existing \gls{O-RAN} interfaces (e.g., \texttt{O1}) and conceptually extends them through the proposed \texttt{RO1}, \texttt{RO2}, and \texttt{RO3}  links to support \gls{RIS}-specific control, their practical realization would require either vendor-specific implementations or future standardization efforts coordinated by the corresponding \gls{O-RAN} Working Groups to ensure interoperability and compliance across different equipment vendors.

\subsection{Quantitative Considerations on Control-Loop Latency and Signaling Overhead}
\label{subsec:DARIO-latency}

The operation of \name{} adheres to the hierarchical control loops defined by the \gls{O-RAN} architecture, where orchestration is performed at the Non-\gls{RT} level and enforcement takes place at Near-\gls{RT} time scales. The feasibility of the proposed framework is governed by the time-critical orchestration loop executed by \name{} at each assignment period of duration $I_{\text{time}}$, which must satisfy the timing constraint
$T_{\text{DARIO}} + T_{\text{signal}} + T_{\text{act}} \le I_{\text{time}}$,
where $T_{\text{DARIO}}$ denotes the execution time of the \name{} optimization process, $T_{\text{signal}}$ accounts for inter-layer signaling latency, and $T_{\text{act}}$ represents the Near-\gls{RT} enforcement latency. 

Traffic and channel-related metrics are generated at fine time granularity within the RAN but are exposed to the Non-\gls{RT} domain through the \texttt{O1} interface in an aggregated and asynchronous manner. These analytics are processed by Non-\gls{RT} \emph{rApps} and reused across multiple assignment periods to derive statistical estimations and predictions. As a result, telemetry acquisition is decoupled from the time-critical orchestration loop and does not lie on its critical timing path.

The assignment period $I_{\text{time}}$ is determined by the dynamics of the propagation environment and the mobility of the \glspl{UE}. For pedestrian mobility scenarios, where the average channel conditions evolve slowly over time, $I_{\text{time}}$ can be assumed to be on the order of a few seconds, e.g., $I_{\text{time}} \approx 2~\text{s}$.

The inter-layer signaling latency $T_{\text{signal}}$ is associated with the dissemination of aggregated orchestration decisions through the proposed \texttt{RO2} and \texttt{RO3} interfaces. These interfaces operate at the same functional level as the \texttt{A1} interface in \gls{O-RAN} and are used for policy and configuration dissemination. The signaling latency is therefore expected to lie in the range $T_{\text{signal}} \in [10,100]~\text{ms}$. The Near-\gls{RT} enforcement latency $T_{\text{act}}$ corresponds to the effective application of orchestration decisions in the RAN. It includes \emph{xApp} execution and enforcement through the \texttt{E2} interface. This process involves computation and synchronization at Near-\gls{RT} time scales. As a result, the actuation latency can be bounded as $T_{\text{act}} \in [30,150]~\text{ms}$.

Considering the most restrictive case within the above ranges, i.e., $T_{\text{signal}} = 100~\text{ms}$ and $T_{\text{act}} = 150~\text{ms}$, and assuming a pedestrian mobility scenario with $I_{\text{time}} \approx 2~\text{s}$, the execution time of the \name{} optimization must satisfy
$T_{\text{DARIO}} \le I_{\text{time}} - (T_{\text{signal}} + T_{\text{act}}) \approx 1.75~\text{s}$.
This inequality defines the execution-time threshold that \name{} must satisfy, which is shown to be met in Section~\ref{subsec:complexity}.

\color{black}

\subsection{Scalability of \name{} in Multi-Cell Environments}\label{subsec:DARIO-scalability}

The current design of \name{} centralizes the coordination of multiple \glspl{RIS} within a single \gls{BS} domain. Extending its operation to multi-cell networks could be achieved by deploying one \name{} instance per cell, each responsible for the \glspl{RIS} and \glspl{UE} served by its \gls{BS}. The orchestration logic would  remain identical, as each instance would perform the same delay and reliability aware optimization within its own coverage area.

Two configurations would be possible for analytics provisioning. In moderately sized urban networks, a single set of Non-\gls{RT} \emph{rApps}, i.e., \emph{Traffic Estimator}, \emph{Channel Capacity Estimator}, \emph{RB Utilization}, would generate per-cell telemetry and feed all \name{} instances through indexed data streams. In larger city-wide or regional deployments, these \emph{rApps} would also be instantiated per zone or per cell to reduce telemetry transport delay and better exploit local monitoring sources. In both cases, each \name{} instance would operate autonomously, without requiring coordination among cells.

The physical placement of \name{} instances may be adapted to the scale and latency constraints of the network. For compact deployments, all instances may run as parallel processes within the same Non-\gls{RT} \gls{RIC}. For large-scale urban networks, instances can be distributed geographically, for example at edge nodes closer to the corresponding \glspl{BS}. At the logical level, the \gls{SMO} and Non-\gls{RT} \gls{RIC} remain single and global, providing unified management and analytics while supporting multiple \name{} instances in parallel.

\section{System Model}\label{sec:system-model}
We consider the \gls{UL} operation of a \gls{SIMO} system in a dense urban environment, providing delay-critical connectivity to a set $\mathcal{U}$ of \glspl{UE}. Each \gls{UE} $u \in \mathcal{U}$ generates packets with arbitrary size and arrival distribution. The \glspl{UE} are equipped with a single antenna, while the \gls{BS} has $N_{ant}$ antennas, operating in the sub-6 GHz band. Within the coverage area, a set $\mathcal{R}$ of \gls{RIS} devices are deployed, each composed of $L$ passive elements. The \gls{RIS} configuration during \textit{scheduling period} $t \in \mathcal{T}_i$ is described by the phase-shift matrix $\mathbf{\Phi}_{r,t} = \mathrm{diag}\left(\kappa_{r,1,t} e^{j\phi_{r,1,t}}, \ldots, \kappa_{r,L,t} e^{j\phi_{r,L,t}}\right)$, where $\kappa_{r,l,t} \in [0,1]$ and $\phi_{r,l,t} \in [0, 2\pi)$ are the amplitude and phase-shift of element $l \in [1,L]$ in \gls{RIS} $r \in \mathcal{R}$. Each \gls{RIS} $r$ has a local \gls{RIS} Controller that applies $\mathbf{\Phi}_{r,t}$ according to the configurations provided by \name{} at the beginning of each \textit{scheduling period} $t \in \mathcal{T}_i$.

Table \ref{tab:notation} summarizes the main notation used in the system model and subsequent analysis.

\begin{table*}[t]
\caption{Summary of notation and parameters.}
\label{tab:notation}
\centering
\renewcommand{\arraystretch}{1.08}
\begin{tabularx}{\textwidth}{@{}p{0.13\textwidth}p{0.37\textwidth}p{0.13\textwidth}p{0.37\textwidth}@{}}
\toprule
\textbf{Symbol} & \textbf{Definition} & \textbf{Symbol} & \textbf{Definition} \\
\midrule
$\mathcal{U}$ & Set of UEs & $\mathcal{R}$ & Set of RISs \\
$u\in\mathcal{U}$ & UE index & $r\in\mathcal{R}$ & RIS index \\
$i\in\mathcal{I}$ & Assignment period index & $t\in\mathcal{T}_i$ & Scheduling period index \\
$I_{\scriptstyle time}$ & Assignment period duration [s] & $T_{\scriptstyle time}$ & Scheduling period duration $=I_{\scriptstyle time}/|\mathcal{T}_i|$ [s] \\
$N_{\mathrm{ant}}$ & BS receive antennas & $L$ & Elements per RIS \\
$\mathbf{\Phi}_{r,t}$ & RIS-$r$ phase-shift matrix at period $t$ & $\kappa_{r,l,t}$ & Amplitude of RIS element $l$ \\
$\phi_{r,l,t}$ & Phase of RIS element $l$ [rad] & $B$ & Phase-quantization bits per element \\
$d$ & Inter-element spacing at RIS [m] & $\theta_0$ & Reflection angle (broadside reference) [rad] \\
$B_{\mathrm{sys}}$ & System bandwidth [Hz] & $N_{\mathrm{sc}}$ & Subcarriers per RB \\
$\Delta_f$ & Subcarrier spacing [Hz] & $OH$ & Overhead fraction for control plane in 5G \\
$N_{\text{cell}}^{\text{RB}}$ & RBs in cell & $N_u^{\text{RB}}$ & RBs assigned to UE $u$ \\
$\pi_{u,i}$ & WRR weight of UE $u$ at period $i$ & $t_{\text{slot}}$ & TTI duration [s] \\
$P_{u,n}^{\text{tx}}$ & UE $u$ TX power on RB $n$ [W] & $N_0$ & Noise PSD [W/Hz] \\
$P_{u,n}^{\text{pl,S1}}$ & Path loss UE–BS in S1 & $P_{u,r,n}^{\text{pl,S2}}$ & Cascaded UE–RIS–BS path loss in S2 \\
$\mathbf{h}_{u,n}^{\text{S1}}$ & UE–BS fast fading in S1 & $\mathbf{g}_{u,r,l,n}$ & RIS-$l$–BS vector channel \\
$h_{u,r,l,n}$ & UE–RIS-$l$ scalar channel & $K_{u,r},K_r$ & Rician factors UE–RIS, RIS–BS \\
$\overline{\gamma}_{u,n}^{\text{S1}}$ & Avg. SNR  in S1 & $\overline{\gamma}_{u,r,n}^{\text{S2}}$ & Avg. SNR in S2 \\
$\gamma_{u,n}^{\text{S1}}$ & Instantaneous SNR in S1 & $\gamma_{u,r,n}^{\text{S2}}$ & Instantaneous SNR in S2 \\
$\eta_m$ & Spectral efficiency of MCS $m$ [bit/s/Hz] & $[\gamma_m^{\min},\gamma_m^{\max}]$ & SNR region of MCS $m$ \\
$p_{u,m}^{\text{S1}}$ & $\Pr\{\text{MCS}=m \mid \text{S1}\}$ & $p_{u,r,m}^{\text{S2}}$ & $\Pr\{\text{MCS}=m \mid \text{S2 with RIS }r\}$ \\
$\omega_u^{\text{S1}}$ & Time fraction in S1 for UE $u$ & $\omega_{u,r}^{\text{S2}}$ & Time fraction in S2 using RIS $r$ \\
$x_{u,r,t}$ & Binary: RIS $r$ assigned to $u$ at $t$ & $G_{\mathrm{loss}}$ & Avg. amplitude reduction \\
$\lambda$ & Noncentrality parameter in S2 & $Q_{N_{\mathrm{ant}}}(\cdot,\cdot)$ & Marcum-$Q$ function \\
$A_u(\tau,t)$ & Arrival process [bit] & $S_u(\tau,t)$ & Service process [bit] \\
$B_u(i)$ & Arrivals at TTI $i$ [bit] & $C_u(j)$ & Served bits at TTI $j$ [bit] \\
$M_{A_u}(\theta)$ & MGF of arrivals & $M_{S_u}(-\theta)$ & Negative MGF of service \\
$\rho_A,\sigma_A$ & EBB rate and burst params. & $\rho_S,\sigma_S$ & EBF rate and burst params. \\
$\varepsilon$ & Delay violation target & $W_u$ & Delay bound of UE $u$ [s] \\
$W_u^{\text{th}}$ & Target delay bound [s] & $\delta,\theta$ & SNC tuning parameters \\
$f_{\text{obj}}$ & Objective function in Eq.~\eqref{eq:ObjectiveFunction} &  &  \\
\bottomrule
\end{tabularx}
\end{table*}

\subsection{Radio Resource Scheduling Model}

The serving \gls{OFDMA} cell has $N_{\text{cell}}^{\text{RB}} = \left\lfloor \frac{B}{N_{\text{sc}}\Delta_f}(1 - OH) \right\rfloor$ \glspl{RB}, where $B$ is bandwidth, $N_{\text{sc}}$ the subcarriers per \gls{RB}, $\Delta_f$ the subcarrier spacing, and $OH$ the control signaling overhead. 
\gls{RB} allocation among active \glspl{UE} follows a \gls{WRR} policy\footnote{Note that O-RAN is assumed to execute this \gls{WRR} policy dynamically in the Near-\gls{RT} \gls{RIC} via an \emph{xApp}, while \name{} operates at the Non-\gls{RT} level and uses this assumption to infer the expected share of radio resources per \gls{UE} for its delay bound and reliability estimations.}, enabling proportional \gls{RB} assignment at each \textit{assignment period} $i\in \mathcal{I}$ based on pre-assigned weights $\pi_{u,i}$, normalized such that $\sum_{u\in \mathcal{U}} \pi_{u,i} = 1$. Each \gls{UE} receives $N_u^{\text{RB}} = N_{\text{cell}}^{\text{RB}} \pi_{u,i}$ \glspl{RB}. While exploring optimal \gls{RB} scheduling algorithms is beyond this paper’s scope, see e.g.,~\cite{Abu-Ali2014}, \gls{WRR} enables differentiation of resource assignments according to \gls{UE}'s delay bound and reliability requirements.

The \gls{BS} adapts the \gls{MCS} per \gls{UE} considering the instantaneous channel quality, yielding spectral efficiencies $\eta_m$~bps/Hz indexed by $m \in [1, N_c]$, where $N_c$ is the total number of available \glspl{MCS}. Each \gls{MCS} is valid over an \gls{SNR} range $[\gamma_m^{min}, \gamma_m^{max}]$. For each \gls{UE} $u \in \mathcal{U}$, $p_{u,m,n}$ denotes the probability of achieving $\eta_m$ in \gls{RB} $n \in [1, N_{\text{cell}}^{\text{RB}}]$ during \textit{scheduling period} $t \in \mathcal{T}_i$, reflecting channel quality and impacting transmission delay including buffering. Note that the values of $\eta_m$, $\gamma_m^{\min}$, and $\gamma_m^{\max}$ follow the 3GPP TS~38.214 specification~\cite{3gpp-ts-38214}. Two scenarios are considered in a \textit{scheduling period} $t \in \mathcal{T}_i$, $\forall i \in \mathcal{I}$: \textit{Scenario 1 ($S1$)}: the \gls{UE} is not assigned to any \gls{RIS} device, due to \gls{RIS} occupancy by other \gls{UE} or lack of \gls{LOS} with any \gls{RIS}. \textit{Scenario 2 ($S2$)}: the \gls{UE} is assigned to a \gls{RIS}, establishing a dynamically reconfigurable \gls{LOS} to the \gls{BS} via the \gls{RIS}.

The following subsections detail the computation of $p_{u,m,n}$ for scenarios \textit{S1} and \textit{S2}. Our analysis assumes the adoption of \gls{ICIC} techniques to mitigate \gls{UL} interference between \glspl{UE} in different cells~\cite{He2018}.

\subsection{Channel Model Scenario 1: Non-\gls{RIS} Assignment}

When \name{} does not assign a \gls{RIS} to \gls{UE}~$u$, \gls{RIS} devices act as passive scatterers. The received flat-fading signal at the \gls{BS} on \gls{RB}~$n$ is $\mathbf{y}_{u,n}^{\text{S1}} = \sqrt{P_{u,n}^{\text{tx}} P_{u,n}^{\text{pl,S1}}} \, \mathbf{h}_{u,n}^{\text{S1}} s_{u,n} + \mathbf{n}_o$, where $s_{u,n}$ is the transmitted symbol, $P_{u,n}^{\text{tx}}$ its power, and $P_{u,n}^{\text{pl,S1}}$ the path loss between \gls{UE}~$u$ and the \gls{BS}. The Rayleigh fading vector $\mathbf{h}_{u,n}^{\text{S1}} \in \mathbb{C}^{N_{\text{ant}} \times 1}$ has i.i.d. entries $\sim \mathcal{CN}(0,1)$, and $\mathbf{n}_o \sim \mathcal{CN}(\mathbf{0}, N_o \mathbf{I}_{N_{\text{ant}}})$ is AWGN.

Assuming the \gls{BS} applies \gls{MRC}, the received vector $\mathbf{y}_{u,n}^{\text{S1}}$ is projected onto the conjugate of the channel vector $\mathbf{h}_{u,n}^{\text{S1}}$, assuming perfect \gls{CSI}\footnote{The accuracy of the CSI depends on the estimation techniques (e.g., \cite{Haochen2024,Cheng2024}) implemented by the Channel Capacity Estimator rApp in the Non-RT RIC. Since CSI is only used to derive long-term service statistics required by the SNC model and is not a decision variable of \name{}, CSI inaccuracies affect the numerical tightness of the estimated delay bounds but do not alter the operation of the DARIO framework nor the relative behavior among compared baseline solutions. Accordingly, our analysis assumes the most favorable case of perfect CSI estimation.} at the \gls{BS}, in order to maximize the output \gls{SNR}. This results in $r_{u,n}^{\text{S1}} = \sqrt{P_{u,n}^{\text{tx}} P_{u,n}^{\text{pl,S1}}} \, \|\mathbf{h}_{u,n}^{\text{S1}}\|^2 s_{u,n} + \mathbf{h}_{u,n}^{\text{S1},H} \mathbf{n}_o$, with instantaneous \gls{SNR} $\gamma_{u,n}^{\text{S1}} = \overline{\gamma}_{u,n}^{\text{S1}} \|\mathbf{h}_{u,n}^{\text{S1}}\|^2$, where $\overline{\gamma}_{u,n}^{\text{S1}} = P_{u,n}^{\text{tx}} P_{u,n}^{\text{pl,S1}}/N_o$.

Given the i.i.d. $\mathcal{CN}(0,1)$ structure of $\mathbf{h}_{u,n}^{\text{S1}}$, the squared norm $\left\| \mathbf{h}_{u,n}^{\text{S1}} \right\|^2$  follows a chi-squared distribution with $2N_{\text{ant}}$ degrees of freedom. Therefore, $\gamma_{u,n}^{\text{S1}}$ follows a Gamma distribution with \gls{PDF}:
\begin{equation}
    f_{\gamma_u}^{\text{S1}}(\gamma) = \frac{1}{\overline{\gamma}_u^{\text{S1}}} \cdot \frac{1}{(N_{\text{ant}} - 1)!} \left( \frac{\gamma}{\overline{\gamma}_u^{\text{S1}}} \right)^{N_{\text{ant}} - 1} \exp\left( -\frac{\gamma}{\overline{\gamma}_u^{\text{S1}}} \right),
\end{equation}

The probability of achieving a target spectral efficiency $\eta_m$, where $m \in [1, N_c]$, is:
\begin{equation}
    p_{u,m}^{\text{S1}} = F_{\Gamma}( \gamma_m^{\text{max}}; N_{\text{ant}}, \overline{\gamma}_u^{\text{S1}} ) - F_{\Gamma}( \gamma_m^{\text{min}}; N_{\text{ant}}, \overline{\gamma}_u^{\text{S1}} ),
\end{equation}
where $F_{\Gamma}(\cdot; k, \theta)$ is the Gamma \gls{CDF} with shape $k = N_{\text{ant}}$ and scale $\theta = \overline{\gamma}_u^{\text{S1}}$.

\subsection{Channel Model Scenario 2: RIS Assignment} 

\name{} assigns the \gls{RIS} $r \in \mathcal{R}$ to the \gls{UE} $u \in \mathcal{U}_{r}$, where $\mathcal{U}_{r}$ is the set of \glspl{UE} with \gls{LOS} to $r$. The remaining \gls{RIS} devices $r' \in \mathcal{R} \setminus \{r\}$ act as passive scatterers and are not explicitly modeled in the channel equations, as their impact is assumed negligible due to random phase alignment. The received signal at the \gls{BS} on \gls{RB} $n$ is $\mathbf{y}_{u,r,n}^{S2} = \sqrt{P_{u,n}^{\mathrm{tx}} P_{u,r,n}^{\mathrm{pl},S2}}\, s_{u,n} \sum_{l=1}^{L} \mathbf{g}_{u,r,l,n} \, \psi_{r,l,n} \, h_{u,r,l,n} + \mathbf{n}_0$, where $P_{u,r,n}^{\mathrm{pl},S2}$ is the path loss over the cascaded \gls{UE}–\gls{RIS}–\gls{BS} link, and $\psi_{r,l,n} = \kappa_{r,l,n} e^{j \phi_{r,l,n}}$ is the reflection coefficient of the $l$-th RIS element, with amplitude $\kappa_{r,l,n} \in [0,1]$ and phase $\phi_{r,l,n} \in [0, 2\pi)$. The term $h_{u,r,l,n} \in \mathbb{C}$ is the fast-fading coefficient from UE $u$ to element $l$ of RIS $r$, and $\mathbf{g}_{u,r,l,n} \in \mathbb{C}^{N_{\mathrm{ant}} \times 1}$ is the vector channel from element $l$ to the BS antennas. Both links follow Rician fading due to LOS conditions~\cite{Sang2024}: $h_{u,r,l,n} = \sqrt{\tfrac{K_{u,r}}{K_{u,r} + 1}} \bar{h}_{u,r,l,n} + \sqrt{\tfrac{1}{K_{u,r} + 1}} \tilde{h}_{u,r,l,n}$, with Rician factor $K_{u,r}$, deterministic LOS component $\bar{h}_{u,r,l,n}$, and scattered component $\tilde{h}_{u,r,l,n} \sim \mathcal{CN}(0,1)$. Similarly, $\mathbf{g}_{u,r,l,n} = \sqrt{\tfrac{K_r}{K_r + 1}} \bar{\mathbf{g}}_{r,l,n} + \sqrt{\tfrac{1}{K_r + 1}} \tilde{\mathbf{g}}_{r,l,n}$, where $K_r$ is the Rician factor, $\bar{\mathbf{g}}_{r,l,n}$ the LOS component, and $\tilde{\mathbf{g}}_{r,l,n} \sim \mathcal{CN}(\mathbf{0}, \mathbf{I}_{N_{\mathrm{ant}}})$ the scattered term. We assume the \gls{RIS} elements introduce uncorrelated scattering in both the \gls{UE}–\gls{RIS} and \gls{RIS}–\gls{BS} links. Accordingly, the components of $\tilde{h}_{u,r,l,n}$ and $\tilde{\mathbf{g}}_{r,l,n}$ are modeled as i.i.d. across elements. This assumption simplifies the statistical characterization of the cascaded channel.

\noindent \textbf{\gls{RIS} Phase Quantization and Frequency Response Effects.}
Each RIS element applies a quantized phase shift $\phi_{r,l,n}$ using a uniform $B$-bit phase shifter~\cite{Xu2021}, constant across all \glspl{RB} assigned to UE $u$. Thus, phase values lie in the set $\phi_{r,l,n} \in \left\{ 0, \Delta \phi, \ldots, (2^B - 1)\Delta \phi \right\}, \quad \Delta \phi = \frac{2\pi}{2^B}$. The \gls{RIS} controller computes a continuous ideal phase $\phi_{r,l,n_0}^{\mathrm{ideal}}$ based on the central RB $n_0$, which is then quantized as $\hat{\phi}_{r,l,n} = \arg \min_{\phi \in \{0, \Delta \phi, \ldots\}} \left| \phi - \phi_{r,l,n_0}^{\mathrm{ideal}} \right|$.

Due to channel frequency selectivity, the ideal phase at RB $n$ differs by a frequency-induced deviation $\xi_{r,l,n}$: $\phi_{r,l,n}^{\mathrm{ideal}} = \phi_{r,l,n_0}^{\mathrm{ideal}} + \xi_{r,l,n}$. The total phase error is then
\begin{equation}
\epsilon_{r,l,n} = \hat{\phi}_{r,l,n} - \phi_{r,l,n}^{\mathrm{ideal}} = \underbrace{\hat{\phi}_{r,l,n} - \phi_{r,l,n_0}^{\mathrm{ideal}}}_{\epsilon_{r,l,n}^{\mathrm{quant}}} - \underbrace{\xi_{r,l,n}}_{\epsilon_{r,l,n}^{\mathrm{freq}}},
\end{equation}
with quantization $\epsilon_{r,l,n}^{\mathrm{quant}}$ and frequency $\epsilon_{r,l,n}^{\mathrm{freq}}$ components assumed independent, as they arise from separate mechanisms: the former due to discrete phase control, and the latter arising from the frequency-dependent phase progression across the RIS aperture, which produces a deterministic spatial phase rotation for subcarriers away from the central frequency $f_c$. These phase errors impair the coherent combining gain of the \gls{RIS}. The expected complex reflection gain for a single element is $\mathbb{E} \left[ e^{j \epsilon_{r,l,n}} \right] = \mathbb{E} \left[ e^{j \epsilon_{r,l,n}^{\mathrm{quant}}} \right] \cdot \mathbb{E} \left[ e^{-j \xi_{r,l,n}} \right]$.

Assuming $\epsilon_{r,l}^{\mathrm{quant}} \sim \mathcal{U}\left[-\frac{\Delta \phi}{2}, \frac{\Delta \phi}{2} \right]$~\cite{Badiu2019}, we get $\mathbb{E} \left[ e^{j \epsilon_{r,l}^{\mathrm{quant}}} \right] = \text{sinc} \left( \Delta \phi/2\right)$. The second term accounts for the deterministic frequency-dependent phase deviation among subcarriers caused by the fixed phase configuration of the RIS at the central frequency $f_c$. For the $m$-th subcarrier, this deviation is
\begin{equation}
\xi_{r,l,m} = \frac{2\pi (f_m - f_c)d \sin\theta_0}{c}\,(l-1),
\end{equation}
where $d$ is the inter-element spacing, $\theta_0$ the reflection angle, $c$ the speed of light, and $l$ the element index along the RIS aperture. The coherent gain over all $12N_u^{\mathrm{RB}}$ subcarriers assigned to the UE is then given by
\begin{equation}
\mathbb{E}\!\left[e^{-j \xi_{r,l,n}}\right]
\approx
\frac{1}{12N_u^{\mathrm{RB}}}
\sum_{m=1}^{12N_u^{\mathrm{RB}}}
\left|
\frac{\sin\!\left(\tfrac{L \varsigma_m}{2}\right)}
     {L \sin\!\left(\tfrac{\varsigma_m}{2}\right)}
\right|,
\end{equation}
where $\varsigma_m = \frac{2\pi (f_m - f_c)d \sin\theta_0}{c}$.

Combining both effects, the average amplitude reduction per RIS element is
\begin{equation}
G_{\mathrm{loss}} =
\text{sinc}\!\left(\frac{\Delta\phi}{2}\right)
\cdot
\frac{1}{12N_u^{\mathrm{RB}}}
\sum_{m=1}^{12N_u^{\mathrm{RB}}}
\left|
\frac{\sin\!\left(\tfrac{L \varsigma_m}{2}\right)}
     {L \sin\!\left(\tfrac{\varsigma_m}{2}\right)}
\right|.
\end{equation}

\noindent \textbf{Received Signal and SNR with Phase Errors.}
The phase errors impair the coherent superposition of the reflected paths. The effective combining vector at the \gls{BS} is $\mathbf{z}_{u,r,n} = \sum_{l=1}^L \mathbf{g}_{u,r,l,n} \, h_{u,r,l,n} \, e^{j \epsilon_{r,l,n}}$, and the received signal vector becomes $\mathbf{y}_{u,r,n}^{S2} = \sqrt{P_{u,n}^{\mathrm{tx}} P_{u,r,n}^{\mathrm{pl},S2}}\, s_{u,n} \mathbf{z}_{u,r,n} + \mathbf{n}_0$, 
and the scalar output after \gls{MRC} combining is $r_{u,r,n} = \mathbf{z}_{u,r,n}^H \mathbf{y}_{u,r,n}^{S2} = \sqrt{P_{u,n}^{\mathrm{tx}} P_{u,r,n}^{\mathrm{pl},S2}}\, s_{u,n} \|\mathbf{z}_{u,r,n}\|^2 + \mathbf{z}_{u,r,n}^H \mathbf{n}_0$.

The instantaneous \gls{SNR} is $\gamma_{u,r,n}^{S2} =\overline{\gamma}_{u,r,n}^{S2} \|\mathbf{z}_{u,r,n}\|^2$, with $\overline{\gamma}_{u,r,n}^{S2} = P_{u,n}^{\mathrm{tx}} P_{u,r,n}^{\mathrm{pl},S2}/N_0$ the average \gls{SNR}.

\noindent \textbf{Probability of Selecting MCS Index.}
Due to the LOS components in both \gls{UE}–\gls{RIS} and \gls{RIS}–\gls{BS} links, $\mathbf{z}_{u,r,n}$ comprises independent Rician-distributed terms, and thus $\|\mathbf{z}_{u,r,n}\|^2$ follows a non-central chi-square distribution with $2N_{\mathrm{ant}}$ degrees of freedom. Its non-centrality parameter is $\lambda = L^2 \cdot \frac{K_{u,r}}{K_{u,r} + 1} \cdot \frac{K_r}{K_r + 1} \cdot G_{\mathrm{loss}}^2$, where $K_{u,r}$ and $K_r$ are the Rician factors. We adopt $K_{u,r} = 3\,\mathrm{dB}$ and $K_r = 6\,\mathrm{dB}$, representing typical partial and stable LOS conditions for \gls{UE}–\gls{RIS} and \gls{RIS}–\gls{BS} paths, respectively.

The \gls{PDF} of $\gamma_{u,r,n}^{S2}$ is
\begin{equation}
\begin{split}
f_{\gamma_{u,r,n}^{S2}}(\gamma) &= \frac{1}{\overline{\gamma}_{u,r,n}^{S2}} \left( \frac{\gamma}{\lambda \overline{\gamma}_{u,r,n}^{S2}} \right)^{\frac{N_{\mathrm{ant}} - 1}{2}} \\
&\quad \cdot \exp\left( -\frac{\gamma + \lambda \overline{\gamma}_{u,r,n}^{S2}}{\overline{\gamma}_{u,r,n}^{S2}} \right) I_{N_{\mathrm{ant}} - 1} \left( \frac{2 \sqrt{\lambda \gamma}}{\overline{\gamma}_{u,r,n}^{S2}} \right),
\end{split}
\end{equation}
where $I_{N_{\mathrm{ant}} - 1}(\cdot)$ is the modified Bessel function of the first kind. The \gls{CDF} is $F_{\gamma_{u,r,n}^{S2}}(\gamma) = 1 - Q_{N_{\mathrm{ant}}} \left( \sqrt{\lambda}, \sqrt{\frac{\gamma}{\overline{\gamma}_{u,r,n}^{S2}}} \right)$, where $Q_{N_{\mathrm{ant}}}(\cdot,\cdot)$ is the Marcum-Q function. Thus, the probability of selecting MCS index $m$, defined over $[\gamma_m^{\min}, \gamma_m^{\max}]$, is
\begin{equation}
\begin{split}
p_{u,m}^{S2} 
&= Q_{N_{\mathrm{ant}}} \left( \sqrt{\lambda}, \sqrt{\frac{\gamma_m^{\min}}{\overline{\gamma}_{u,r,n}^{S2}}} \right) - Q_{N_{\mathrm{ant}}} \left( \sqrt{\lambda}, \sqrt{\frac{\gamma_m^{\max}}{\overline{\gamma}_{u,r,n}^{S2}}} \right).
\end{split}
\end{equation}

\section{Per-UE Delay Bound Modeling via Stochastic Network Calculus}\label{sec:SNCmodel}

This section derives the per-\gls{UE} delay bound $W$ using the \gls{SNC} framework for a multi-\gls{RIS} deployment scenario under dynamic traffic and channel conditions. We briefly revisit the fundamental principles of \gls{SNC} and the procedure to derive arrival and service envelopes and the corresponding delay bound. The traffic model is characterized from empirical observations, while the service process is formulated from the cell capacity under dynamic \gls{RIS} operation. In contrast to conventional \gls{SNC} formulations assuming a stationary and ergodic service process, the proposed analysis extends the framework to configuration-dependent and time-varying conditions induced by \gls{RIS} reconfiguration, by modeling the service process as an explicit function of UE--RIS assignment decisions. This leads to a finite-mixture service characterization whose parameters evolve with the scheduling configuration, enabling direct analytical evaluation of how different RIS configurations reshape the resulting delay bound.

\subsection{Introduction to Stochastic Network Calculus (SNC)}
\gls{SNC} provides an analytical framework for modelling and analyzing the performance of mobile networks, characterized by stochastic packet arrivals and service rates. Let $A(\tau,t)$ and $S(\tau,t)$ denote the cumulative arrival and service processes over the interval $(\tau,t]$, respectively, representing the total number of bits that arrive to, and can be served by, the node during this time. \gls{SNC} uses the concepts of \gls{EBB} and  \gls{EBF}~\cite{Fidler1} to define an upper bound $\alpha (\tau,t)$, known as arrival envelope, and a lower bound, $\beta  (\tau,t)$, known as service envelope, as specified in the following
\begin{equation}
   \text{P}[A(\tau,t) > \alpha(\tau,t)] \leq \varepsilon_{ A}, \, \text{P}\left[ S(\tau,t) < \beta(\tau,t) \right] \leq \varepsilon_{ S},
    \label{eq:EBBAffine}
\end{equation}
where $\varepsilon_A$ and $\varepsilon_S$ are the overflow and deficit profiles~\cite{Fidler1}. We assume affine functions for  $\alpha (\tau,t)$ and $\beta  (\tau,t)$  which are defined in Eqs.~\eqref{eq:AffineArrivalEnvelope} and \eqref{eq:ServiceEnvelopeBeta}. The parameters $\rho_{ A} > 0$, $\rho_{ S} > 0$ and $b_{ A} \geq 0$, $b_{ S} \geq 0$ are the rate and burst parameters for these functions. Additionally, $\left[x \right]_{+}$ denotes $\text{max}\{0,x \}$. Finally, $\delta> 0$ is a sample path argument considered by \gls{EBB} and \gls{EBF} models~\cite{Fidler1}.
\begin{equation}
    \alpha (\tau,t) = (\rho_{ A}+\delta)[t-\tau] + b_{ A},
    \label{eq:AffineArrivalEnvelope}
\end{equation}
\begin{equation}
    \beta (\tau,t)= (\rho_{ S} -\delta)\left[t-\tau -\sfrac{b_{ S}}{\rho_{ S}} \right]_{+}.
    \label{eq:ServiceEnvelopeBeta}
\end{equation}

Representing $\alpha(\tau,t)$ and $\beta(\tau,t)$ over a $t-\tau$ axis, we can obtain the delay bound $W$ as the horizontal deviation between these envelopes, as shown in Eq.~\eqref{eq:Delay_bound_v2}. This deviation occurs when the slope of the service envelope is greater than the slope of the arrival envelope, i.e., $\frac{\partial \beta(\tau, t)}{\partial (t - \tau)} > \frac{\partial \alpha(\tau, t)}{\partial (t - \tau)} $, provided that the stability condition $\rho_{ S} - \delta  > \rho_{ A} + \delta$ is satisfied, ensuring that the delay bound remains finite.
\begin{equation}
    W = \tfrac{b_{ A} + b_{ S}}{\rho_{ S}-\delta}.
    \label{eq:Delay_bound_v2}
\end{equation}

For further details on \gls{SNC} fundamentals, the reader is referred to~\cite{leboudec2001network,jiang2008stochastic}.

\subsection{Methodology for Delay Bound Derivation  with SNC}
We summarize the generic methodology  introduced in ~\cite[Sections II.A and II.B]{Fidler1} to derive the arrival and service envelopes, and subsequently compute the corresponding delay bound:

\textit{1)} Compute the \glspl{MGF} for the arrival and service processes, i.e., $M_{ A}(\theta)$ and $M_{ S}(\smallminus\theta)$. The \gls{MGF} of a process $X$ is $\text{E}\left[e^{\theta X}\right]$ with a tunable parameter $\theta$. 

\textit{2)} Define upper bounds for the \glspl{MGF} as shown in Eqs.~\eqref{eq:ArrivalAffineEnvelopeModel} and \eqref{eq:AffineEnvelopeServiceMGF}. They are characterized by the rate parameters $\rho_{ A}$ and $\rho_{ S}$, and the burst parameters $\sigma_{ A}$ and $\sigma_{ S}$. Note that $\rho_{ A}$ and $\rho_{ S}$ match those in Eqs.~\eqref{eq:AffineArrivalEnvelope} and \eqref{eq:ServiceEnvelopeBeta}. Fit the \glspl{MGF} computed in Step 1 to the exponential bounds defined in Eq.~\eqref{eq:ArrivalAffineEnvelopeModel} and \eqref{eq:AffineEnvelopeServiceMGF}  to extract the parameters $\rho_{ A}$, $\sigma_{ A}$, $\rho_{ S}$, and $\sigma_{ S}$.
\begin{equation}
        M_{ A}(\theta) \leq \text{e}^{ \theta \left(\rho_{ A}[t-\tau] + \sigma_{ A} \right)},
        \label{eq:ArrivalAffineEnvelopeModel}
    \end{equation}
    \begin{equation}
        M_{ S}(\smallminus\theta) \leq \text{e}^{ \smallminus\theta \left(\rho_{ S}[t-\tau] -\sigma_{ S} \right)}.
        \label{eq:AffineEnvelopeServiceMGF}
    \end{equation}

\textit{3)} The \gls{EBB} and \gls{EBF} models defined in Eq.~\eqref{eq:EBBAffine}, and the \glspl{MGF} are directly connected by the Chernoff bound~\cite{ross2014first}. Based on this, we can obtain $b_{ A}$ and $b_{ S}$ as shown in Eqs.~\eqref{eq:Burst_parameter_arrival} and \eqref{eq:burst_parameter_service}. We equally distribute the target violation probability among $\varepsilon_{ A}$ and $\varepsilon_{ S}$, i.e., $\varepsilon_{ A}=\varepsilon_{ S}=\varepsilon/2$.
\begin{equation}
        b_{ A} = \sigma_{ A} - \tfrac{1}{\theta} \left[\text{ln}(\varepsilon_{ A}) + \text{ln}\left( 1-\text{e}^{\smallminus\theta \delta} \right) \right],
        \label{eq:Burst_parameter_arrival}
    \end{equation}
    \begin{equation}
        b_{ S} = \sigma_{ S} - \tfrac{1}{\theta} \left[\text{ln}(\varepsilon_{ S}) + \text{ln}\left( 1-\text{e}^{\smallminus\theta \delta }\right) \right].
        \label{eq:burst_parameter_service}
    \end{equation}

\textit{4)} Combining the expressions for $b_{ A}$ and $b_{ S}$  with the delay bound expression in Eq.~\eqref{eq:Delay_bound_v2}, the delay bound $W$ can be reformulated as:
\begin{equation}
    W = \frac{ \sigma_{ A}+\sigma_{ S}-\tfrac{2}{\theta}\left[ \text{ln} \left(\frac{\varepsilon}{2} \right) + \text{ln} \left(1-\text{exp}\left[-\theta\delta\right] \right)\right] }{\rho_{ S} -\delta}.
    \label{eq:Middle_delay_bound}
\end{equation} 

\gls{SNC} models typically yield conservative estimations of the delay bound~\cite{Fidler1}. Hence, it is necessary to optimize the tuneable parameters  $\theta$ and $\delta$ to minimize  $W$ and obtain a bound as tight as possible.

\subsection{UE Traffic Model for the Arrival Process}\label{sec:uRLLCTrafficModel}
In a slotted system with \gls{TTI} duration $t_{ slot}$, the cumulative arrival process $A_{ u}(\tau,t)$ for \gls{UE} $u$ is given by

\begin{equation}
    A_{ u}(\tau,t)=\sum_{i=\lceil\tau/t_{slot}\rceil+1}^{\lfloor t/t_{slot}\rfloor} B_{ u}(i),
    \label{eq:ArrivalProcesUE_real}
\end{equation}
where $B_{ u}(i)$ represents the number of bits that arrive to the transmission buffer in the $i$-th \gls{TTI}.

In this work, we consider that each \gls{UE} may generate traffic following an arbitrary distribution. For this reason, we assume that the \gls{PMF} of $B_{ u}(i)$ can be estimated by using samples of the incoming bits per \gls{TTI} in the last $T_{ OBS}$ \glspl{TTI}. Specifically, we define the sample vector $\vec{x}_{ B_{ u}} =\{b_{ u,1}, b_{ u,2}\, \hdots\, b_{ u,T_{ OBS}} \}$, where $b_{ u,i}$ denotes the number of bits that arrive to the transmission buffer in the \gls{TTI} $i$ for the \gls{UE} $u$. We assume the elements of $\vec{x}_{ B_{ u}}$ are independent and identically distributed random variables, taken from a sliding window of the most recent $T_{OBS}$ \glspl{TTI}.  Note that in our prior work~\cite{Adamuz2024}, we demonstrated that  $T_{ OBS}\geq4000$ is required to accurately estimate the \gls{PMF} of the generated traffic. Additionally, $b_{ u,i}=\sum_{j=1}^{J_{ u,i}}l_j$, where $J_{ u,i}$ is the number of packets generated by the \gls{UE} $u$ in the \gls{TTI} $i$ and $l_{ j}$ the size of the packet $j$. Note that the computation of $\vec{x}_{ B_{ u}}$ is a task performed by the Traffic Estimator \emph{rApp}.

With these assumptions, the empirical \gls{MGF} of $B_{ u}(i)$ over the observation window is given by
\begin{equation}
    M_{ B_{ u}}(\theta) =\frac{1}{T_{ OBS}} \sum_{i=1}^{T_{ OBS}} \text{exp}\left[\theta b_{ u,i}\right].
    \label{eq:MGFTraffic_real}
\end{equation}

Since the samples of $B_u(i)$ are i.i.d., the \gls{MGF} of the arrival process $A_{ u}(\tau,t)$ is $ M_{ A_{ u}}(\theta) = \text{E}\left[M_{ B_{ u}}(\theta)^{N_{ slot}(\tau,t)} \right]$, being $N_{slot}(\tau,t) =  \left\lfloor t/t_{\text{slot}} \right\rfloor - \left\lceil \tau/t_{\text{slot}} \right\rceil$ the number of \glspl{TTI} fully contained in the interval $(\tau, t]$.  By exponentiating the logarithm of the expectation's argument for $M_{ A_{ u}}(\theta)$, we get $ M_{ A_{ u}}(\nu) = \text{E}\left[\text{e}^{\nu N_{ slot}(\tau,t)} \right]$, with $v=\text{ln}\left[M_{ B_{ u}}(\theta)\right]$. Since $N_{ slot}(\tau,t)$ is a constant, $ M_{ A_{ u}}(\nu) = \text{e}^{\nu(t-\tau)/t_{ slot}} $. We can replace $\nu$ and get
\begin{equation}
            M_{ A_{ u}}(\theta)   = e^{\left[\text{ln}\left( \tfrac{1}{T_{ OBS}} \sum_{i=1}^{T_{ OBS}} \text{exp}\left[\theta b_{ u,i}\right]\right) \tfrac{ (t-\tau)}{t_{ slot}}\right]}.
    \label{eq:MGFTraffic2_real}
\end{equation}

Finally, by equaling the left and right sides of Eq. \eqref{eq:ArrivalAffineEnvelopeModel}, we obtain $\rho_{ A_{ u}}(\theta)$ as Eq. \ref{Eq:ArrivalEnvelopeParameters_real} shows. Note that $\sigma_{ A_{ u}}(\theta)=0$.
\begin{equation}
        \rho_{ A_{ u}}(\theta) = \frac{\text{ln}\left[\tfrac{1}{T_{ OBS}} \sum_{i=1}^{T_{ OBS}} \text{exp}\left[\theta b_{ u,i}\right]\right] }{\theta t_{ slot}}.  
\label{Eq:ArrivalEnvelopeParameters_real}
\end{equation}

\subsection{Cell Capacity Model for the Service Process}\label{sec:ServiceModel}
The service process $S_{ u}(\tau,t)$ represents the accumulated capacity provided to \gls{UE} $u$ and is defined by Eq.~\eqref{eq:ServiceCurveRANSlice}. $C_{ u}(j)$ is the number of bits served by the cell to \gls{UE} $u$ in \gls{TTI} $j$, as specified in Eq.~\eqref{eq:CapacityTTIue}. Additionally, $\eta_{u,n}(j)$ denotes the spectral efficiency (bps/Hz) for \gls{UE} $u$ in \gls{TTI} $j$ on \gls{RB} $n$. We assume $\eta_{u,n}$ is an i.i.d. random variable across \glspl{RB} and \glspl{TTI}, which simplifies the derivation. Although in practice, spectral efficiency may exhibit spatial and temporal correlation, this assumption enables tractable \gls{MGF}-based analysis under \gls{SNC}.
\begin{equation}
S_{ u}(\tau,t) = \sum_{j=\lceil\tau/t_{slot}\rceil+1}^{\lfloor t/t_{slot}\rfloor} C_{ u}(j),
    \label{eq:ServiceCurveRANSlice}
\end{equation} 
\begin{equation}
     C_{ u}(j) = \sum_{ n = 1}^{ N_{ u}^{ RB} } N_{ sc}\Delta_{ f}t_{ slot} \eta_{u,n}(j).
     \label{eq:CapacityTTIue}
\end{equation}

The negative \gls{MGF} for $\eta_{u,n}(j)$ is defined in Eq.~\eqref{eq:MGFAvailableCapacityforUE}. It depends on the probabilities $\omega_{u}^{S1}$ and $\omega_{u,r}^{S2}$, $\forall r \in \mathcal{R}_u$, that characterize the fraction of scheduling periods in which \gls{UE} $u$ operates without \gls{RIS} assistance or is served through a specific \gls{RIS}, respectively, together with the corresponding \gls{MCS} spectral efficiencies $\eta_m$ and selection probabilities $p_{u,m}^{S1}$ and $p_{u,r,m}^{S2}$.  This formulation yields a finite-mixture service characterization in which the mixture weights $\omega_{u}^{S1}$ and $\omega_{u,r}^{S2}$ are set by the UE--RIS configuration, while \gls{CSI} only affects the MCS selection probabilities $p_{ u,r,m}^{ S2}$ and $p_{ u,m}^{ S1}$.
\begin{equation}
\begin{split}
        M_{ \eta}(\smallminus \theta) =& \sum_{ r\in \mathcal{R}_{ u}}\omega_{ u,r}^{ S2}\sum_{ m=1}^{ N_{ c}}\text{exp}\left(\smallminus \theta \eta_m\right) p_{ u,r,m}^{ S2}\\
        & + \omega_{ u}^{ S1}\sum_{ m=1}^{ N_{ c}}\text{exp}\left(\smallminus \theta \eta_m\right) p_{ u,m}^{ S1}.
\end{split}
    \label{eq:MGFAvailableCapacityforUE}
\end{equation}

The parameter $\omega_{ u,r}^{ S2} = \left(\sum_{ t \in \mathcal{T}_i} x_{ u,r,t} \right) /   |\mathcal{T}_i|$ is the probability  \gls{UE} $u$ being assigned a specific \gls{RIS} $ r\in \mathcal{R}_{ u}$ during the \textit{scheduling period} $t\in \mathcal{T}_{ i}$, and $x_{ u,r,t} \in \{0,1\}$ is a scheduling decision variable indicating whether \gls{RIS} $r$ is assigned to UE $u$ at \textit{scheduling period} $t$.  Finally, $\omega_{ u}^{ S1} = 1-\sum_{ r \in \mathcal{R}_{ u}} \omega_{ u,r}^{ S2}$.  Note that Section~\ref{sec:ProblemFormulation} details how $x_{ u,r,t}$ is set to minimize the packet transmission delay for all \glspl{UE}.
The \gls{MGF} of $C_{ u}(j)$ is defined in Eq.~\eqref{eq:DemostrationMGF_C}. We represent $M_{ C}(\smallminus \theta)$ in terms of $ M_{ \eta}(\smallminus \theta)$, leveraging on the \gls{MGF} properties related to linear combination of independent random variables. Next, by substituting  $\text{ln}\left( M_{\eta}(\smallminus  N_{ sc} \Delta_{ f}t_{ slot} \theta)\right)= \nu$, we obtain the \gls{MGF} of the process $N_{ u}^{ RB}$ as a function of the free parameter $\nu$. 

\begin{equation}
    \begin{split} 
    M_{ C}(\smallminus \theta) & = 
        \text{E}\left[\left[M_{\eta}(\smallminus  N_{ sc} \Delta_{ f}t_{ slot} \theta)\right]^{N_{ u}^{ RB}}\right] \\
        & =   E\left[e^{\nu N_{ u}^{ RB}} \right] = M_{ N_{ u}^{ RB}} (\nu).
    \end{split}
     \label{eq:DemostrationMGF_C}
\end{equation}

Since $M_{ N_{ u}^{ RB}} (\nu) = \text{e}^{N_{ u}^{ RB}\nu}$, we can define $M_{ C_u}(\smallminus \theta)$ as follows
\begin{equation}
\begin{split}
        M_{ C_u}(\smallminus \theta) &= \left(\sum_{ r\in \mathcal{R}_{ u}}\omega_{ u,r}^{ S2}\sum_{ m=1}^{ N_{ c}}\text{e}^{\left(\smallminus \theta N_{ sc} \Delta_{ f}t_{ slot} \eta_m\right)} \right.\\
        & p_{ u,r,m}^{ S2} \left. + \omega_{ u}^{ S1}\sum_{ m=1}^{ N_{ c}}\text{e}^{\left(\smallminus \theta N_{ sc} \Delta_{ f}t_{ slot} \eta_m\right)}p_{ u,m}^{ S1}\right)^{N_{ u}^{ RB}}.
\end{split}\label{eq: MGF_C}
\end{equation}
Considering $M_{ C_{ u}}(\smallminus\theta)$, we can compute the negative \gls{MGF} of the service process $S_{ u}(\tau,t)$ as the following
            $M_{ S_{ u}}(\smallminus\theta)  = \text{e}^{ \text{ln}\left( M_{ C_{ u}}(\smallminus\theta) \right) \frac{(t-\tau)}{t_{slot}}}$.
    \label{eq:MGFranslice1}
%
Finally, by equaling the left and right sides of Eq.~\eqref{eq:AffineEnvelopeServiceMGF}, we obtain $\rho_{ S_{ u}}(\theta)$ as follows. Note that $\sigma_{ S_{ u}}(\theta)=0$.
\begin{equation}
        \rho_{ S_{ u}}(\theta) =  \frac{-\text{ln}\left[ M_{ C_u}(\smallminus \theta) \right]}{\theta t_{ slot}}.
\label{Eq:ServiceEnvelopeParameters}
\end{equation}

\subsection{Delay Bound Estimation for Packet Transmission }\label{sec:DelayBoundEstimation}
Using the results from Sections \ref{sec:uRLLCTrafficModel} and \ref{sec:ServiceModel}, we can define the delay bound $W_{ u}$ as a function of the tuneable parameters $\theta$ and $\delta$, as specified in Eq.~ \eqref{eq:Definite_delay_bound}. To estimate $W_{ u}$, we need to solve the following optimization problem:\\
\noindent \textbf{Problem}~\texttt{DELAY\_BOUND\_PACKET\_TRANSMISSION}:
\begin{alignat}{2}
&\underset{\theta,\delta}{\text{min}}  \quad && W_u= \frac{2 t_{ slot} \left[ \text{ln} \left(\frac{\varepsilon_{ u}}{2} \right) + \text{ln} \left(1-\text{exp} \left[\smallminus \theta\delta\right ] \right)\right] }{- \theta t_{ slot}\rho_{ S_{ u}}(\theta)  + \delta \theta t_{ slot}}.\label{eq:Definite_delay_bound}\\
&\text{s.t.: } \quad && \theta,\delta,\rho_{ A_{ u}}(\theta),\rho_{ S_{ u}}(\theta) >0, \label{eq:ProblemFormulationCons1Delay}\\
& \quad && \rho_{ S_{ u}}(\theta) - \rho_{ A_{ u}}(\theta) > 2\delta. \label{eq:ProblemFormulationCons4}
\end{alignat}

The above-mentioned problem involves a non-convex objective function within a non-convex region, leading to multiple local minima and making exhaustive search computationally infeasible. To address the non-convexity, we adapt a heuristic inspired by~\cite{Adamuz2024}. However, unlike~\cite{Adamuz2024}, which focuses on non-\gls{RIS} systems, our method supports dynamic multi-\gls{RIS} configurations and mobile \glspl{UE}. These aspects introduce new constraints and variability in the delay model that are not considered in prior work.

\section{UE-RIS Assignment Problem Formulation}\label{sec:ProblemFormulation}

For each \textit{assignment period} $i \in \mathcal{I}$, \name{} determines in advance a one-to-one \gls{UE}-\gls{RIS} assignment for every \textit{scheduling period} $t \in \mathcal{T}_{i}$. The \texttt{UE-RIS-RB\_ASSIGNMENT} problem is formulated as follows.

\noindent \textbf{Problem}~\texttt{UE-RIS-RB\_ASSIGNMENT}:
\begin{alignat}{2}
&\underset{x_{u,r,t},\,N_{u}^{RB}}{\text{min}} \quad &&  
f_{obj} =
\max_{u \in \mathcal{U} \setminus \mathcal{U}_{\mathcal{R}}} \left\{ \frac{W_{u}[\boldsymbol{\omega_{u}}]}{W_{u}^{th}}\right\} +
\max_{u \in \mathcal{U}_{\mathcal{R}}}\left\{ \frac{W_{u}[\boldsymbol{\omega_{u}}]}{W_{u}^{th}}\right\},
\label{eq:ObjectiveFunction} \\
&\text{s.t.: }  \quad && \sum_{t \in \mathcal{T}_{i}} \sum_{r \notin \mathcal{R}_{u}} x_{u,r,t} = 0, \quad \forall u \in \mathcal{U}_{\mathcal{R}}, \label{eq:ProblemFormulationCons0}\\
& \quad && \sum_{r \in \mathcal{R}} x_{u,r,t} \leq 1, \quad \forall t \in \mathcal{T}_{i}, \quad \forall u \in \mathcal{U}_{\mathcal{R}}, \label{eq:ProblemFormulationCons1}\\
& \quad && \sum_{u \in \mathcal{U}_{\mathcal{R}}} x_{u,r,t} \leq 1, \quad \forall t \in \mathcal{T}_{i}, \quad \forall r \in \mathcal{R}, \label{eq:ProblemFormulationCons2}\\
& \quad && \sum_{u \in \mathcal{U}} N_{u}^{RB} = N_{cell}^{RB}. \label{eq:ProblemFormulationCons3}
\end{alignat}

In this problem, Eq.~\eqref{eq:ObjectiveFunction} minimizes the sum of two worst-case delay ratios: the first term considers the maximum ratio of the experienced delay bound $W_{u}$ to the target delay bound $W_{u}^{th}$ over the \glspl{UE} without \gls{LOS} to any \gls{RIS} (i.e., $u \in \mathcal{U} \setminus \mathcal{U}_{\mathcal{R}}$), while the second term considers the maximum ratio over the subset $\mathcal{U}_{\mathcal{R}}$ of \glspl{UE} with \gls{LOS} to at least one \gls{RIS}. Note that $W_{u}$ is computed with respect to a target violation probability $\varepsilon_{u}$ $\forall u \in \mathcal{U}$. Focusing on a single \gls{UE} $u \in \mathcal{U}$, the experienced delay bound $W_{u}\!\left[\mathbf{\omega_{u}}\right]$ depends on the probability vector $\mathbf{\omega_{u}} = \{\omega_{u}^{S1}, \omega_{u,r}^{S2}\}$ $\forall r \in \mathcal{R}$, as defined by Eqs.~\eqref{eq: MGF_C}, \eqref{Eq:ServiceEnvelopeParameters} and \eqref{eq:Definite_delay_bound}. This dependence is explicit and analytical, enabling the optimization variables to directly influence the SNC service process through the configuration-dependent mixture weights. In turn, these probabilities depend on the binary and integer decision variables $x_{u,r,t}$ and $N_{u}^{RB}$, respectively, which are constrained by the following: Constraint~\eqref{eq:ProblemFormulationCons0} enforces that any \gls{UE} $u$ lacking \gls{LOS} to \gls{RIS} $r$, i.e., $r \notin \mathcal{R}_{u}$, cannot be assigned that \gls{RIS}; Constraint~\eqref{eq:ProblemFormulationCons1} enforces that one \gls{RIS} is assigned to \gls{UE} $u$ in a \textit{scheduling period} $t \in \mathcal{T}_{i}$; Constraint~\eqref{eq:ProblemFormulationCons2} enforces that each \gls{RIS} $r \in \mathcal{R}$ serves a single \gls{UE} in a \textit{scheduling period} $t \in \mathcal{T}_{i}$; and Constraint~\eqref{eq:ProblemFormulationCons3} enforces that the total number of \glspl{RB} allocated across all \glspl{UE} equals the available \glspl{RB} in the cell.

The objective function in Eq.~\eqref{eq:ObjectiveFunction} depends on $W_{ u}$ $\forall u \in \mathcal{U}$, and involves a non-linear relationship with the weights $\left\{\omega_{ u}^{ S1},\omega_{ u,r}^{ S2}\right\}$ $\forall r \in \mathcal{R}$, and in turn with the decision variables $x_{u,r,t}$ $\forall r \in \mathcal{R}$, $\forall t \in \mathcal{T}_{ i}$ and the integer RB allocation variables $N_u^{RB}$ $\forall u \in \mathcal{U}$, as shown in Eqs.~\eqref{eq: MGF_C}, \eqref{Eq:ServiceEnvelopeParameters} and \eqref{eq:Definite_delay_bound},  making the \texttt{UE-RIS-RB\_ASSIGNMENT} problem a \gls{NIP} problem~\cite{Hemmecke2010}. Given (a) the nonlinearity of the objective function, (b) the large number of decision variables, and (c) the need to compute the delay bound $W_{u}$ $\forall u \in \mathcal{U}$ by invoking a heuristic (see Section~\ref{sec:DelayBoundEstimation}) for every possible \gls{UE}-\gls{RIS} assignment (exhaustive search) at each \textit{scheduling period} $t \in \mathcal{T}_{i}$, the problem becomes computationally prohibitive. The total number of \gls{UE}-\gls{RIS} assignment combinations is $|\mathcal{C}|^{|\mathcal{T}_{i}|}$, where $|\mathcal{C}|$ is the number of possible assignments within a single \textit{scheduling period}. Because \gls{RB} allocation to \glspl{UE} is performed once and remains fixed across all \textit{scheduling periods} $\mathcal{T}_i$, the total number of possible \gls{RB} allocation is $\binom{N_{\text{cell}}^{\mathrm{RB}} - 1}{|\mathcal{U}| - 1}$. Therefore, the total number of feasible joint configurations is $|\mathcal{C}|^{|\mathcal{T}_i|} \times \binom{N_{\text{cell}}^{\mathrm{RB}} - 1}{|\mathcal{U}| - 1}$, highlighting the exponential growth and computational intractability of exact optimization.

To reduce computational complexity, we propose a two-stage heuristic: \textit{(1)} \gls{RB} allocation among \glspl{UE} assuming an equiprobable \gls{UE}-\gls{RIS} assignment (i.e., no optimized RIS association), as performed in Algorithm~\ref{alg:RB-allocation}; and \textit{(2)} dynamic \gls{UE}-\gls{RIS} association based on the fixed \gls{RB} allocation obtained in the first stage, aiming to improve delay performance. These stages correspond to Algorithm~\ref{alg:RB-allocation} and Algorithm~\ref{alg:Optimum-RIS-UE-Assignment}, respectively, described below.

\begin{algorithm}[t!]

\SetAlgoLined
\small
\textbf{Inputs:} $W_{ u}^{ th}$, $\forall u \in \mathcal{U}$, initial RB allocation $N_u^{RB}= N_{cell}^{RB}/|\mathcal{U}|$\;
\textbf{Initialization:} $f_{obj}^{(old)} = \infty$, $f_{obj}^{(new)} = \infty$, $Cont=False$\;
Compute $W_u$ and $f_{obj}^{(old)} =$ Eq.~\eqref{eq:ObjectiveFunction}\;

\While{\textbf{not} $Cont$}{
    Select $u^{\max} = \arg\max_{u \in \mathcal{U}} \left( W_u/W_u^{th} \right)$\;
    Select $u^{\min} = \arg\min_{u \in \mathcal{U} \; \text{s.t.} \; N_u^{RB} > 1} \left( W_u/W_u^{th} \right)$\;

        Update RB allocation: $N_{u^{\max}}^{RB} = N_{u^{\max}}^{RB} + 1$ and 
        $N_{u^{\min}}^{RB} = N_{u^{\min}}^{RB} - 1$\;
        Recompute $W_u$ and $f_{obj}^{(new)}$\;
        \eIf{$f_{obj}^{(new)} < f_{obj}^{(old)}$}{
            $f_{obj}^{(old)} = f_{obj}^{(new)}$\;
        }{
            Revert the last RB exchange: 
            $N_{u^{\max}}^{RB} = N_{u^{\max}}^{RB} - 1$ and
            $N_{u^{\min}}^{RB} \gets N_{u^{\min}}^{RB} + 1$\;
            $Cont = True$\;
        }
}
\textbf{return:} $N_u^{RB}$\;
\caption{RB allocation with equiprobable UE-RIS assignment}
\label{alg:RB-allocation}

\end{algorithm}

\textbf{Algorithm~\ref{alg:RB-allocation}.} It iteratively redistributes \glspl{RB} among \glspl{UE} to improve the global delay performance. At each step, the \gls{UE} with the highest normalized delay ratio receives one \gls{RB} from the \gls{UE} with the lowest ratio, as long as the donor \gls{UE} retains at least one \gls{RB} (lines 5-7). The objective function is updated after each exchange (line 8), and the process continues until no further improvement is observed (lines 9-13).

Algorithm~\ref{alg:RB-allocation} has a worst-case complexity of $\mathcal{O}(N_{\text{cell}}^{RB} \cdot |\mathcal{U}|)$. Each iteration scans all \glspl{UE} to find $u^{\max}$ and $u^{\min}$, and recomputes the objective function with cost $\mathcal{O}(|\mathcal{U}|)$. Only \gls{RB} exchanges reducing the objective are accepted; since the objective decreases discretely and is bounded below, convergence occurs after a finite number of steps. Conservatively, we upper bound iterations by $N_{\text{cell}}^{RB}$ for this complexity estimation.
\begin{algorithm}[t!]
\SetAlgoLined
\small
\textbf{Inputs:} $W_{ u}^{ th}$, $\varepsilon_{ u}$  $\forall u \in \mathcal{U}_{ \mathcal{R}}$\; 
\textbf{Initialization:} $\mathcal{U}_{eva}= \emptyset$, $\mathcal{U}_{ don} = \emptyset$ $f_{ obj}^{ (old)}=\infty$, $f_{ obj}^{ (new)}=\infty$, $Cont=False$\;
Set $x_{r,t,u}$ and $x_{r,t,u}^{ \prime} = x_{r,t,u}$\;
Compute $\mathbf{\omega_{ u}}$, $W_{ u}$ $\forall u \in \mathcal{U}_{ \mathcal{R}}$, and $f_{ obj}^{ (old)}$\;
 \While{\textbf{not} $Cont$}{
 Select UE $u^{\prime} \in \mathcal{U}_{ \mathcal{R}} \backslash \mathcal{U}_{ eva}$ with maximum $W_{ u^{\prime}} / W_{ u^{\prime}}^{ th}$\;
 Select UE $u^{\prime\prime} \in \mathcal{U}_{ \mathcal{R}^{ u^{\prime}}} \backslash \left(\mathcal{U}_{ eva} \cup \mathcal{U}_{ don} \right) $ with minimum $W_{ u^{\prime\prime}} / W_{ u^{\prime\prime}}^{ th}$\;
 \eIf{$u^{ \prime \prime} == \emptyset$}{$\mathcal{U}_{ eva} = \mathcal{U}_{ eva} \cup u^{ \prime}$}{
 Set $\mathcal{R}_{ dis} = \emptyset$ and $stop = False$\;
\While{$\mathcal{R}_{ dis} \neq \mathcal{R}_{ u^{\prime} } \cup \mathcal{R}_{ u^{\prime \prime}} $ \textbf{or} $stop==False$}{
Set $r = \arg\min\{d(r^{ \prime },u^{ \prime})\} $ $ \forall r^{ \prime}  \in \mathcal{R}_{ u^{\prime} } \cup \mathcal{R}_{ u^{\prime \prime}} \backslash  \mathcal{R}_{ dis}$\;
Randomly select $t \in \mathcal{T}_{ i, u^{\prime \prime},r}$\;
\eIf{$t \neq \emptyset$}{Set $x_{ u^{\prime},r,t}^{ \prime}=1$ and $x_{ u^{\prime\prime},r,t}^{ \prime}=0$\;
Compute $W_{ u^{\prime}}^{\prime}$, $W_{ u^{\prime\prime}}^{\prime}$, and $f_{ obj}^{(new)}$\;
\eIf{$f_{ obj}^{(new)} \leq f_{ obj}^{(old)}$}{Update $x_{u,r,t} = x_{u,r,t}^{ \prime}$, $W_{u} = W_{u}^{\prime}$, and $f_{ obj}^{(old)} = f_{ obj}^{(new)}$\;}{Set $\mathcal{U}_{ don} = \mathcal{U}_{ don} \cup u^{ \prime \prime } $\;}
Set  $stop = True$ \;
}{$\mathcal{R}_{ dis} = \mathcal{R}_{ dis} \cup r$\;}
}
 }
\If{$\mathcal{U}_{ eva}==\mathcal{U}_{\mathcal{R}}$ }{$Cont = True$\;}
 }
 \textbf{return:} $x_{ u,r,t}$, $W_{ u}$ and $f_{ obj}^{(old)}$, \;
 \caption{RIS scheduling}
 \label{alg:Optimum-RIS-UE-Assignment}
\end{algorithm}

\textbf{Algorithm \ref{alg:Optimum-RIS-UE-Assignment}.} It takes as input target delay bounds $W_{ u}^{ th}$ and violation probabilities $\varepsilon_{ u}$ $\forall u \in \mathcal{U}_{ \mathcal{R}}$ (line 1) and initializes the state variables (line 2). It begins with a random \gls{UE}-\gls{RIS} assignment (line 3), and it continues with the calculation of the weight vector $\mathbf{\omega_{ u}}$ and the delay bound $W_{ u}$ for each \gls{UE} $u \in \mathcal{U}_{\mathcal{R}}$, determining the initial objective function value (line 4). The algorithm then iteratively aims to reduce the objective function (lines 5-32).  In each iteration, the \gls{UE} $u^{ \prime}$ with the highest ratio $W_{ u^{\prime}}/ W_{ u^{\prime}}^{ th}$ is selected (line 6), along with the \gls{UE} $u^{ \prime \prime}$ with the lowest ratio $W_{ u^{\prime \prime}}/ W_{ u^{\prime \prime }}^{ th}$ among those experiencing \gls{LOS} with at least one common \gls{RIS} device (line 7). The aim is for  \gls{UE} $u^{\prime}$ to use a common \gls{RIS} $r$ during a \textit{scheduling period} $t \in \mathcal{T}_{ i}$ previously assigned to the \gls{UE} $u^{ \prime \prime}$. If no such \glspl{UE} exist (lines 8-9), \gls{UE} $u^{ \prime}$ is removed from the set of \glspl{UE} that can improve their ratio $W_{ u}/W_{ u}^{ th}$.
If a suitable \gls{UE} exists, another iterative loop starts (lines 12-28), selecting common \gls{RIS} devices, preferentially \gls{RIS} devices closest to $u^{ \prime}$ (line 13). Once a \gls{RIS} $r$ is chosen, a \textit{scheduling period} $t \in \mathcal{T}_{ i}$ is sought for $u^{ \prime}$ to use it (line 14). If such a period exists, the exchange is made (line 16), weights and delay bounds are updated, and the new objective function value is computed (line 17). If the objective function improves, the new \gls{UE}-\gls{RIS} assignment is updated (line 19). If not, \gls{UE} $u^{ \prime \prime}$ is discarded, and the next iteration (outer loop) will attempt another \gls{UE}.
If no suitable \textit{scheduling period} $t \in \mathcal{T}_{ i}$ is found, the nested loop exits (line 21), and another \gls{RIS} $ r  \in \mathcal{R}_{ u^{\prime} } \cup \mathcal{R}_{ u^{\prime \prime}} \backslash  \mathcal{R}_{ dis}$ is considered (line 25). This process repeats until no \glspl{UE} $u^{\prime}$ can improve the ratio between the experienced delay bound and the target one (line 30). Finally, the algorithm returns the \gls{UE}-\gls{RIS} assignment with the lowest objective function (line 33).

The worst-case complexity of Algorithm~\ref{alg:Optimum-RIS-UE-Assignment} is $\mathcal{O}(|\mathcal{U}_{\mathcal{R}}|^2 + |\mathcal{U}_{\mathcal{R}}| \cdot |\mathcal{R}|)$. Each outer iteration selects the \gls{UE} with the highest normalized delay ratio in $\mathcal{O}(|\mathcal{U}_{\mathcal{R}}|)$, then finds a \gls{UE} with shared \gls{LOS} \gls{RIS} also in $\mathcal{O}(|\mathcal{U}_{\mathcal{R}}|)$. The inner loop explores at most $2|\mathcal{R}|$ \gls{RIS} devices, performing objective evaluations with cost $\mathcal{O}(|\mathcal{R}|)$ per iteration. With at most $|\mathcal{U}_{\mathcal{R}}|$ outer iterations, total complexity is as stated.

The computational complexities derived for the proposed algorithms refer to a single \name{} instance operating within one base-station domain. In large-scale deployments, multiple \name{} instances can be deployed in parallel, as described in Section~II.D. Each instance executes the same two-stage optimization procedure independently, preserving the same per-instance complexity. The total computing resources required therefore scale approximately linearly with the number of cells, while the algorithmic complexity of each instance remains unchanged.

\section{Performance Evaluation}
\label{sec:perf_eval}
We evaluate \name{} using a two-stage approach: first, we perform simulations with synthetic traces to rigorously analyze its performance under controlled conditions, then, we validate these results through real-world experiments described in Section~\ref{sec:real}.

In the simulation phase, we model a dense urban environment spanning $250 \times 250 \, \mathrm{m}^2$, featuring a single macro cell~\cite{3gpp-ts-38901} and multiple \gls{RIS} devices. The \gls{BS} operates at $4.7$ GHz (Band n79) with $100$ MHz bandwidth and $60$ KHz subcarrier spacing, providing $135$ available \glspl{RB} to serve delay-critical traffic. Each simulated \gls{RIS} consists of antenna elements arranged in a $10 \times 10$ rectangular grid~\cite{Rossanese2024}.

Simulated scenarios include a varying number of \gls{RIS} devices and \glspl{UE}, where \glspl{UE} move throughout the urban area following the Manhattan mobility model~\cite{Manhattanmodel}. Each \gls{UE} $u \in \mathcal{U}$ generates traffic packets following a Poisson distribution with mean $\lambda_{ u}$, and packet sizes $d_{ u}$ are drawn from an arbitrary distribution. These traffic characteristics are defined analogously to the model presented in Section~\ref{sec:uRLLCTrafficModel}. Additionally, \glspl{UE} are assigned specific delay requirements $W_{ u}^{ th}$ and violation probabilities $\varepsilon_{ u}$.

Simulations were performed using a Python-based simulator on a computing platform equipped with 16 GB RAM and a quad-core Intel Core i7-7700HQ @ 2.80 GHz processor. Table~\ref{tab:ParameterConfig} summarizes the key parameters of the simulation setup. 

\begin{table}[t!]
\centering
\caption{Simulator Parameters Configuration}
\label{tab:ParameterConfig}
\resizebox{\columnwidth}{!}{
\begin{tabular}{|cc|cc|}
\hline
\multicolumn{1}{|c|}{\textbf{Parameters}}                                                    & \textbf{Value}                                             & \multicolumn{1}{c|}{\textbf{Parameters}}                                                                 & \textbf{Value} \\ \hline
\multicolumn{2}{|c|}{\textbf{Cell Layout}}                                                                                                                &  \multicolumn{1}{c|}{\begin{tabular}[c]{@{}c@{}}  Assignment\\ Period $I_{ time}$\end{tabular}}                                                      & 2 s                                                                                            \\ \hline
\multicolumn{1}{|c|}{Size}                                                                   & 250x250 $m^{ 2}$~\cite{3gpp-ts-38901}                                             &  \multicolumn{2}{c|}{\textbf{UE config}}           \\ \hline
\multicolumn{2}{|c|}{\textbf{5G NR \gls{BS}}}                                                                                                         &  \multicolumn{1}{c|}{\begin{tabular}[c]{@{}c@{}}Number of\\ UEs ($|\mathcal{U}|$)\end{tabular}}                                                                       &    $\left\{15,20,25,30, 35\right\}$           \\ \hline
\multicolumn{1}{|c|}{\begin{tabular}[c]{@{}c@{}} Number of \\ antennas  ($N_{ant}$) \end{tabular}} & 8  &  \multicolumn{1}{c|}{UE height}                                                                           &    1.8 m           \\ \hline
\multicolumn{1}{|c|}{Frequency band}                                                         & \begin{tabular}[c]{@{}c@{}}n79 band\\ (central frequency \\ 4.7 GHz)\end{tabular} &  \multicolumn{1}{c|}{\begin{tabular}[c]{@{}c@{}}UE Mobility \\ Model\end{tabular}}                        &    \begin{tabular}[c]{@{}c@{}}Manhattan~\cite{Manhattanmodel}. Speed\\ $v_{ u} \in \left[1,2\right]$ m/s\end{tabular}                                                                           \\ \hline
\multicolumn{1}{|c|}{Bandwidth}                                                              & \begin{tabular}[c]{@{}c@{}}100 MHz\\ (135 RBs)\end{tabular} &  \multicolumn{2}{c|}{\textbf{UE Traffic Pattern}}           \\ \hline
\multicolumn{1}{|c|}{\begin{tabular}[c]{@{}c@{}} Subcarriers\\ per RB ($N_{ sc}$)\end{tabular}} &                                          12                  &  \multicolumn{1}{c|}{\begin{tabular}[c]{@{}c@{}}Traffic\\ Distribution\end{tabular}}                      &   \begin{tabular}[c]{@{}c@{}}Poisson dist. Average rate\\  $\lambda_{ u} \in \left[450,550\right]$ packets/s\end{tabular}         \\ \hline
\multicolumn{1}{|c|}{\begin{tabular}[c]{@{}c@{}} Subcarrier\\ spacing ($\Delta_{ f}$)\end{tabular}}                                                    &       60 KHz                                                     &  \multicolumn{1}{c|}{Packet Size }                                                                         &     \begin{tabular}[c]{@{}c@{}} $\left\{64,128,256,512,1024\right\}$\\ bytes. Uniform Dist.\end{tabular}         \\ \hline
    \gls{BS} height                                                                                                      &          25 m                                                 &  \multicolumn{2}{c|}{\textbf{uRLLC Requirements}}                                                                           \\ \hline
 \multicolumn{1}{|c|}{MAC scheduler}  &   Round robin                                                         & \multicolumn{1}{c|}{Target Delay Bound}                                                                  & \begin{tabular}[c]{@{}c@{}}$\left\{5, 10,15,20,25,\right.$ \\ $\left.50, 100\right\}$ ms\end{tabular}      \\ \hline
                                                        \multicolumn{2}{|c|}{\textbf{RIS}}                                                  &     \multicolumn{1}{c|}{Violation Probability}                                                               &       $\left\{10^{ -3},10^{ -4},10^{ -5}\right\}$     \\ \hline
\multicolumn{1}{|c|}{\begin{tabular}[c]{@{}c@{}}Number of RIS \\ $\left(|\mathcal{R}|\right)$\end{tabular}}                                                         &  \begin{tabular}[c]{@{}c@{}}$\left\{5, 10,15,20,25,\right.$ \\ $\left.30\right\}$\end{tabular}                                                                                                                        &  \multicolumn{2}{c|}{\textbf{Channel Parameters}}                                                                                    \\ \hline
\multicolumn{1}{|c|}{\begin{tabular}[c]{@{}c@{}}RIS elements \\ ($L$)\end{tabular}}                                                       &     100                                                           &  \multicolumn{1}{c|}{\begin{tabular}[c]{@{}c@{}}UE transmission \\ power $\left(P_{ u}^{ tx}\right)$\end{tabular}}                                                               &   24 dBm       \\ \hline
\multicolumn{1}{|c|}{\begin{tabular}[c]{@{}c@{}}RIS height, i.e., \\ placement \end{tabular}}      &    3 m                                                    &  \multicolumn{1}{c|}{\begin{tabular}[c]{@{}c@{}}Thermal noise\\ power\end{tabular}}                                                                 &    -174 dBm/Hz             \\ \hline
\multicolumn{1}{|c|}{\begin{tabular}[c]{@{}c@{}} Number of phase \\ bits ($B$) \end{tabular}} & 3 &  \multicolumn{1}{c|}{\begin{tabular}[c]{@{}c@{}}Loss Propagation\\ model\end{tabular}}                                                                 &  \begin{tabular}[c]{@{}c@{}} 3GPP UMa model\\ (TR 38.901 V18.0.0)~\cite{3gpp-ts-38901}\end{tabular}                \\ \hline
\multicolumn{2}{|c|}{\textbf{DARIO Conf.}} & \multicolumn{1}{c|}{\begin{tabular}[c]{@{}c@{}}Spectral efficiency \\ per MCS $\left(\eta_{ u,k}\right)$\end{tabular}}              & \begin{tabular}[c]{@{}c@{}}See 3GPP  TS 38.214 \\ Table 5.2.2.1-3~\cite{3gpp-ts-38214} \end{tabular}   \\ \hline
\multicolumn{1}{|c|}{\begin{tabular}[c]{@{}c@{}} Scheduling\\ Period $T_{ time}$\end{tabular}}                                                      & 100 ms                                                   & \multicolumn{1}{c|}{\begin{tabular}[c]{@{}c@{}}SNR range per MCS \\ $\left(\left[\gamma_{ k}^{ min}, \gamma_{ k}^{ max}\right]\right)$\end{tabular}} &     \begin{tabular}[c]{@{}c@{}}$\left[2^{\eta_{ u,k}/0.6}-1, \right.$ \\ $\left. 2^{\eta_{ u,k+1}/0.6}-1\right]$ \end{tabular}          \\ \hline
\end{tabular}
}
\end{table}

\subsection{SNC Model Validation under Poisson Traffic}
In the first experiment, we validate the proposed \gls{SNC}-based model for estimating the delay bound $W_{ u}$ of an \gls{UE} $u \in \mathcal{U}$, generating packets with an average rate of $\lambda_u = 2000$ packets/s, across four Validation Scenarios (VS). In VS1, we vary the number of \glspl{RB} allocated to the \gls{UE}, accounting for different distances from the \gls{UE} to the \gls{BS}, and considering both cases with and without exclusive \gls{RIS} allocation, i.e., $\omega_{u,r}^{S2} = 1$. For all cases, we consider a target violation probability $ \varepsilon_{ u} = 10^{ -3} $. In VS2, we fixed the number of \glspl{RB} at $N_{ u}^{ RB}=5$ and vary the violation probability for the previous cases. In VS3, we kept fixed $N_{ u}^{ RB}=5$ and $ \varepsilon_{ u} = 10^{ -3} $, and analyze how $W_{ u}$ varies with different \gls{UE} positions, considering both the presence and absence of a \gls{RIS} device. In VS4, we fixed two \gls{UE} positions and varied the probability of \gls{RIS} assignment $\omega_{u,r}^{ S2}$ within an \textit{assignment period}  $i\in \mathcal{I}$.

\begin{figure}[b!]
    \centering
    \includegraphics[width=\columnwidth]{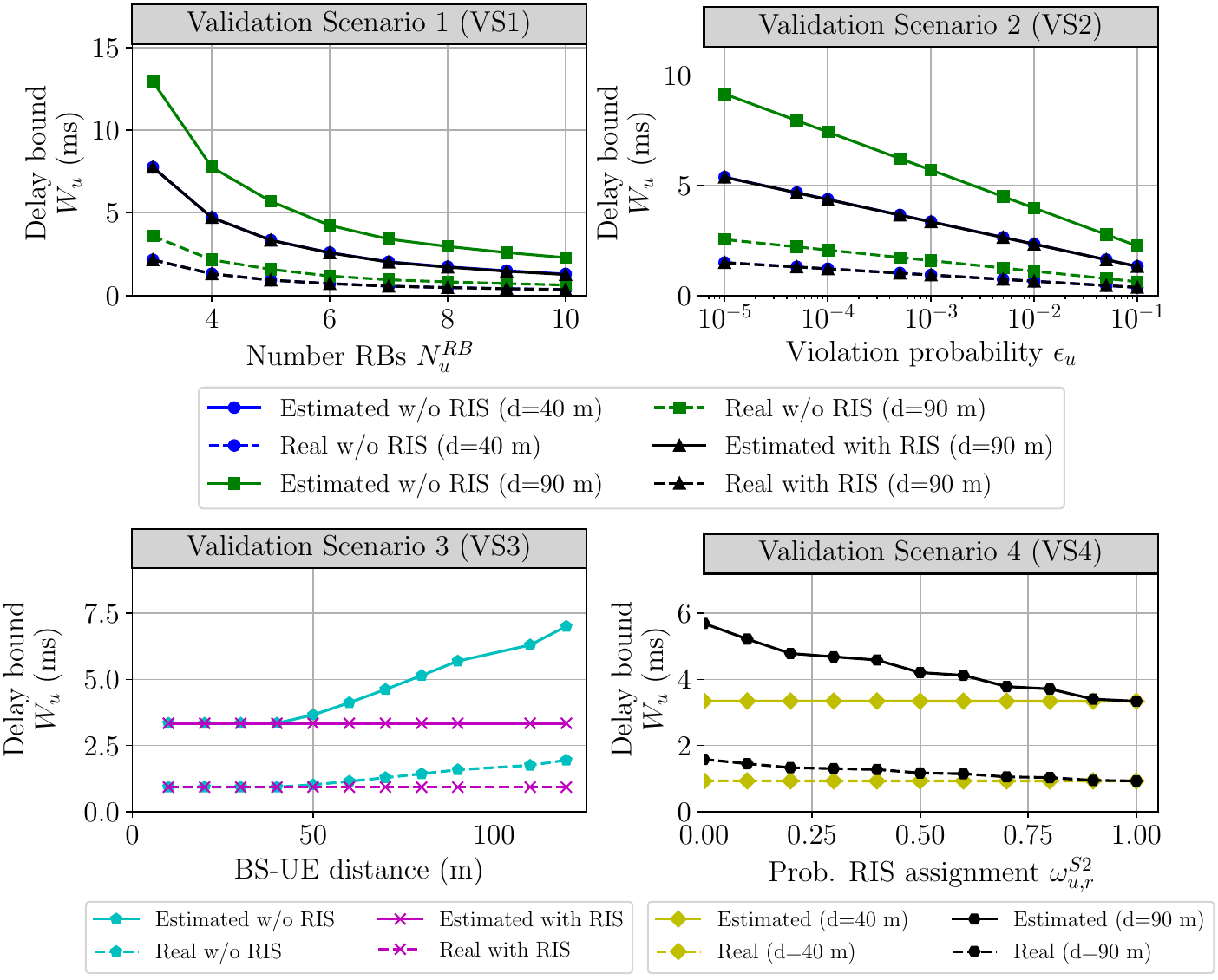}
    \caption{Estimated vs. real delay bounds across multiple validation scenarios (VS). VS1: varying number of RBs, UE distances, and RIS assignment; VS2: fixed RBs, varying violation probability; VS3: fixed RBs and violation probability, varying UE positions with/without RIS; VS4: fixed UE positions, varying RIS assignment probability.}
    \label{fig:snc-validation}
\end{figure}

Fig. \ref{fig:snc-validation} shows the \gls{SNC} model consistently offers a conservative estimation of $ W_{ u} $. Specifically, the \gls{SNC} model yields a relative overestimation close to 230\%, which aligns with the conservative nature of \gls{SNC} modeling frameworks~\cite{Fidler1}. One of the main goals of \gls{SNC}-based models is to account for complex arrival and service processes. In our study, the arrival process is constructed from empirical traffic traces collected at each \gls{TTI}, capturing realistic burstiness patterns. Furthermore the service process is defined in Eq.~\eqref{eq: MGF_C} and is based on a probabilistic model. This model captures the \gls{UL} capacity a \gls{UE} may obtain depending on the dynamic configuration of \gls{RIS} devices, which are managed by \name{} to establish temporal \gls{LOS} links between these \gls{RIS} devices and the \gls{UE}. Considering arrival and service processes that are closer to real-world scenarios justifies the use of \gls{SNC}, which, rather than aiming for an exact match between model and simulator results, provides an upper-bound, conservative estimation of $ W_{ u} $~\cite{Fidler1}. This conservative estimation is ideal for resource planning, as it ensures the fulfillment of \glspl{UE}' delay-critical requirements in real-world scenarios.

The analysis of the different validation scenarios reveals that assigning a \gls{RIS} to an \gls{UE} significantly reduces its experienced delay bound $W_{ u}$. In VS1 and VS2, a \gls{UE} using a \gls{RIS} device at $90$m from the \gls{BS} experiences better delay bound than a \gls{UE} without \gls{RIS} at $40$m from the \gls{BS}. In VS3, the use of a \gls{RIS} device maintains the estimated delay bound nearly invariant across varying \gls{UE}-\gls{BS} distances, outperforming scenarios without \gls{RIS} even when the \gls{UE} is close to the \gls{BS}. In VS4, the delay bound increases as the probability of using a \gls{RIS} within an assignment  period decreases, with significant effects at greater distances from the \gls{BS}.

\subsection{Delay Bound Sensitivity to RIS Phase Quantization}

We perform a sensitivity analysis to evaluate how the quantization of \gls{RIS} phase shifts affects the delay bound $W_u$. The configuration follows the experimental setup described at the beginning of this section, except for the geometric layout and the distance range. The \gls{BS} is placed at $(0,0)$ m, a single \gls{RIS} is located at $(0,500)\,\mathrm{m}$, and one \gls{UE} moves along the $x$-axis from $0$ to $4\,\mathrm{km}$. The \gls{UE} is assigned 5 \glspl{RB} per slot, and the delay bound is computed for a violation probability of $10^{-3}$. This setup allows examining how the delay bound varies with the \gls{RIS} phase quantization level under different propagation conditions.
\begin{figure}[b!]
    \centering
    \includegraphics[width=\columnwidth]{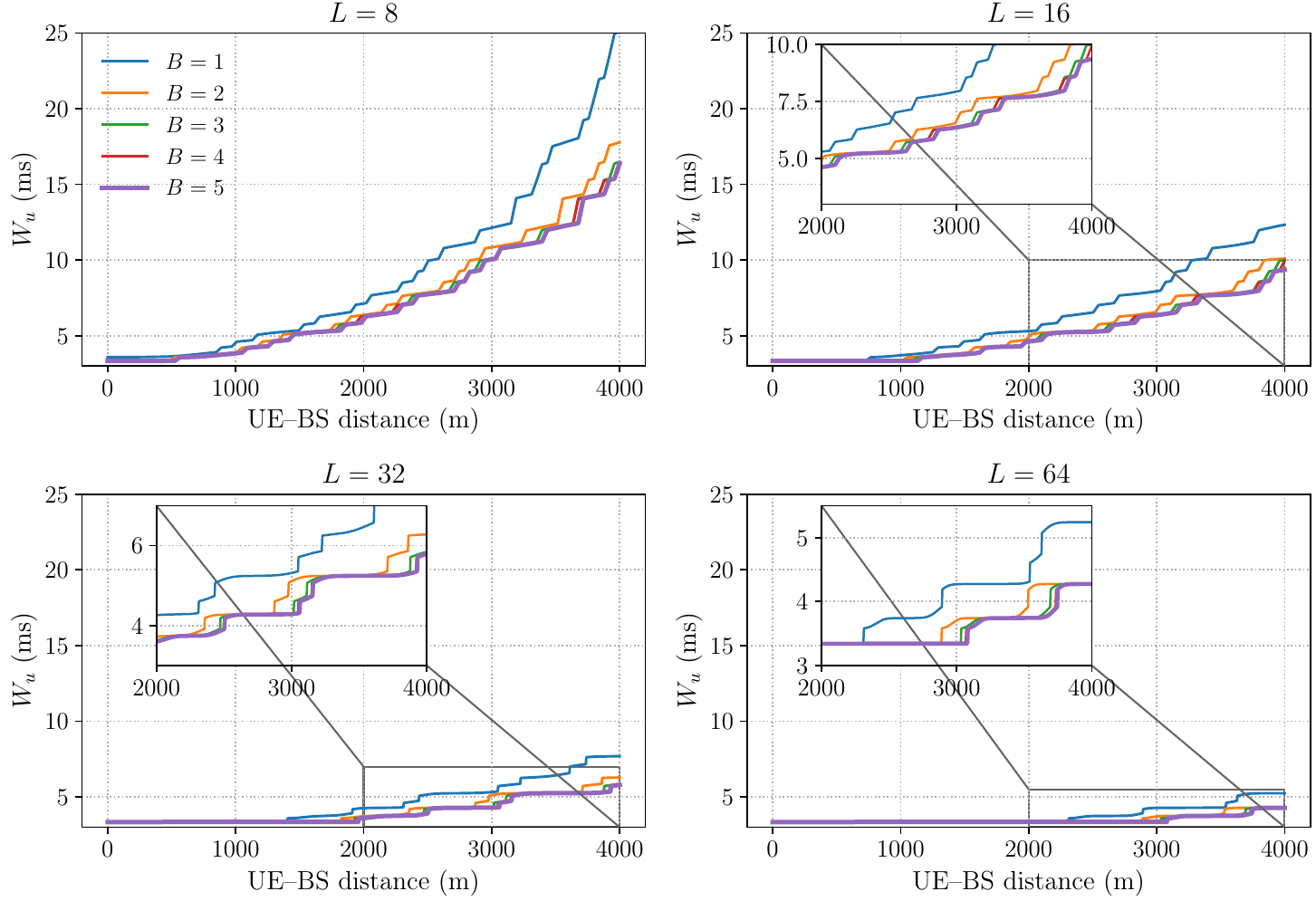}
    \caption{Delay bound $W_u$ versus UE--BS distance for different \gls{RIS} phase quantization levels $B$ and numbers of reflecting elements $L$ ($\varepsilon_u=10^{-3}$, $N_u^{RB}=$ 5~\glspl{RB} per slot).}
    \label{fig:sensivity}
\end{figure}

Figure~\ref{fig:sensivity} depicts the evolution of the delay bound $W_u$ as a function of the \gls{UE}--\gls{BS} distance for different quantization resolutions $B=\{1,2,3,4,5\}$ and numbers of \gls{RIS} elements $L=\{8,16,32,64\}$. The results show that $W_u$ increases exponentially with distance. For small \gls{RIS} configurations such as $L=8$, coarse quantization (few discrete phase states) leads to higher values of $W_u$, since the residual phase error weakens the coherent gain of the reflected components. As $L$ increases, the \gls{RIS} gain compensates for these errors and the delay bound decreases. Moreover, differences between quantization levels become progressively smaller, and from $B=3$ bits onward (eight discrete phase states) the quantization error becomes negligible. This confirms that $B=3$ provides a suitable operating point for \gls{RIS} devices.

In this scenario, the \gls{UE}, the \gls{RIS}, and the \gls{BS} are in mutual \gls{LOS}, which makes the deterministic \gls{LOS} component dominant in the overall channel response. As a result, the sensitivity of the delay bound $W_u$ to the quantization level is substantially reduced. For distances below approximately $1\,\mathrm{km}$ when $L=8$ and up to $2\,\mathrm{km}$ when $L=64$, the effect of the phase resolution on $W_u$ is minimal, since the constructive gain provided by LOS propagation compensates for the discretization error. Therefore, in typical urban deployments with a high density of \gls{RIS} devices placed close to the \glspl{UE} and with clear visibility to both the \gls{BS} and the served \glspl{UE}, coarse phase quantization has a negligible impact on the achievable delay bound.

\subsection{Evaluating \name{} Against the Optimum}\label{sec:HeuristicValidation}
In this experiment, we assess the heuristic algorithms used by \name{}, measuring its deviation from the optimal solution. Due to the problem's complexity, we benchmark our approach against a brute-force algorithm that evaluates all the possible candidate \gls{UE}-\gls{RIS} and \glspl{RB} assignments. For the sake of comprehensibility, we consider a small-scale scenario consisting of  $|\mathcal{R}|=2$ \gls{RIS} devices and $|\mathcal{U}|=6$ \glspl{UE}. In this setup, 2 \glspl{UE} have \gls{LOS} with only one \gls{RIS}, another 2 \glspl{UE} have \gls{LOS} with the other \gls{RIS}, and the remaining 2 \glspl{UE} have \gls{LOS} with both \gls{RIS} devices. This leads to a total of $|\mathcal{C}|=23$ possible \gls{UE}-\gls{UE} assignment combinations in a single \textit{scheduling period} $t \in \mathcal{\mathcal{T}}_{ i}$. Furthermore, we consider $N_{cell}^{RB}\in\{9,10\}$ which leads $\binom{N_{\text{cell}}^{\mathrm{RB}} - 1}{|\mathcal{U}| - 1}$ \gls{RB} allocations, and therefore $23^{|\mathcal{T}_i|}\times\binom{N_{\text{cell}}^{\mathrm{RB}} - 1}{|\mathcal{U}| - 1}$ combinations over $|\mathcal{T}_i|$ \textit{scheduling periods}. 
We evaluate both methods considering different numbers of \textit{scheduling periods} $|\mathcal{T}_{ i}|$ and available \glspl{RB}. The results, summarized in Table~\ref{tab:HerVal}, indicate that in the worst case the optimal value obtained via the brute force approach and the suboptimal solution from the proposed heuristic exhibit a relative error of $9.19$\%.  
\begin{table}[t!]
\centering
\caption{Optimality Gap Analysis of \name{}}
\label{tab:HerVal}
\resizebox{\columnwidth}{!}{
\begin{tabular}{|c|c|c|c|c|c|}
\hline
\begin{tabular}[c]{@{}c@{}}\textbf{$|\mathcal{T}_{ i}|$} / \\ \textbf{$N_{cell}^{RB}$} \end{tabular}           & \textbf{Method} & \textbf{Obj. Func.} & \begin{tabular}[c]{@{}c@{}}\textbf{Relative}\\ \textbf{Error (\%)} \end{tabular}            & \textbf{Iterations} & \textbf{Execution Time} \\ \hline
\multirow{2}{*}{\textbf{1} / \textbf{9}} & \name{}           &  3.0623                   & \multirow{2}{*}{9.1848}  & 9                                   & 178.52 ms                             \\ \cline{2-3} \cline{5-6} 
                            & Brute Force     & 2.8046                   &                          & 1288                                    & 51.21 s                             \\ \hline
\multirow{2}{*}{\textbf{2} / \textbf{9}} & \name{}           & 1.996                   & \multirow{2}{*}{2.0154} & 10                                   & 121.64 ms                            \\ \cline{2-3} \cline{5-6} 
                           & Brute Force     & 1.9568                   &                          & 29624                                  & 16 min. 39 s                             \\ \hline
\multirow{2}{*}{\textbf{3} / \textbf{9}} & \name{}           & 1.8847                   & \multirow{2}{*}{1.0547}  & 11                                   & 126.69 ms                            \\ \cline{2-3} \cline{5-6} 
                            & Brute Force     & 1.8650                  &                          & 681352                                 & 6 h. 33 min. 25 s                            \\ \hline

\multirow{2}{*}{\textbf{1} / \textbf{10}} & \name{}           & 2.0701                   & \multirow{2}{*}{8.7060}  & 7                                   & 101.73 ms                             \\ \cline{2-3} \cline{5-6} 
                            & Brute Force     & 1.9043                   &                          & 2898                                    & 1 min. 39 s                             \\ \hline
\multirow{2}{*}{\textbf{2} / \textbf{10}} & \name{}           & 1.7546                   & \multirow{2}{*}{5.3681} & 10                                   & 112.69 ms                             \\ \cline{2-3} \cline{5-6} 
                           & Brute Force     & 1.6652                   &                          & 66654                                  & 37 min. 35 s                             \\ \hline
\multirow{2}{*}{\textbf{3} / \textbf{10}} & \name{}           & 1.5954                   & \multirow{2}{*}{4.7488}  & 11                                   & 126.69 ms                            \\ \cline{2-3} \cline{5-6} 
                            & Brute Force     & 1.5231                  &                          & 1533042                                 & 14 h. 45 min. 52 s                            \\ \hline
\end{tabular}
}
\end{table}
Such a discrepancy becomes even more reasonable when considering that the brute force method takes nearly $15$ hours to execute for $|\mathcal{T}_{ i}|=3$ and $N_{cell}^{RB}=10$, while \name{} only requires $<180$ ms.
Details about \name's computational complexity are discussed in Section~\ref{subsec:complexity}.

\begin{figure}[t!]
    \centering
    \includegraphics[width=\columnwidth]{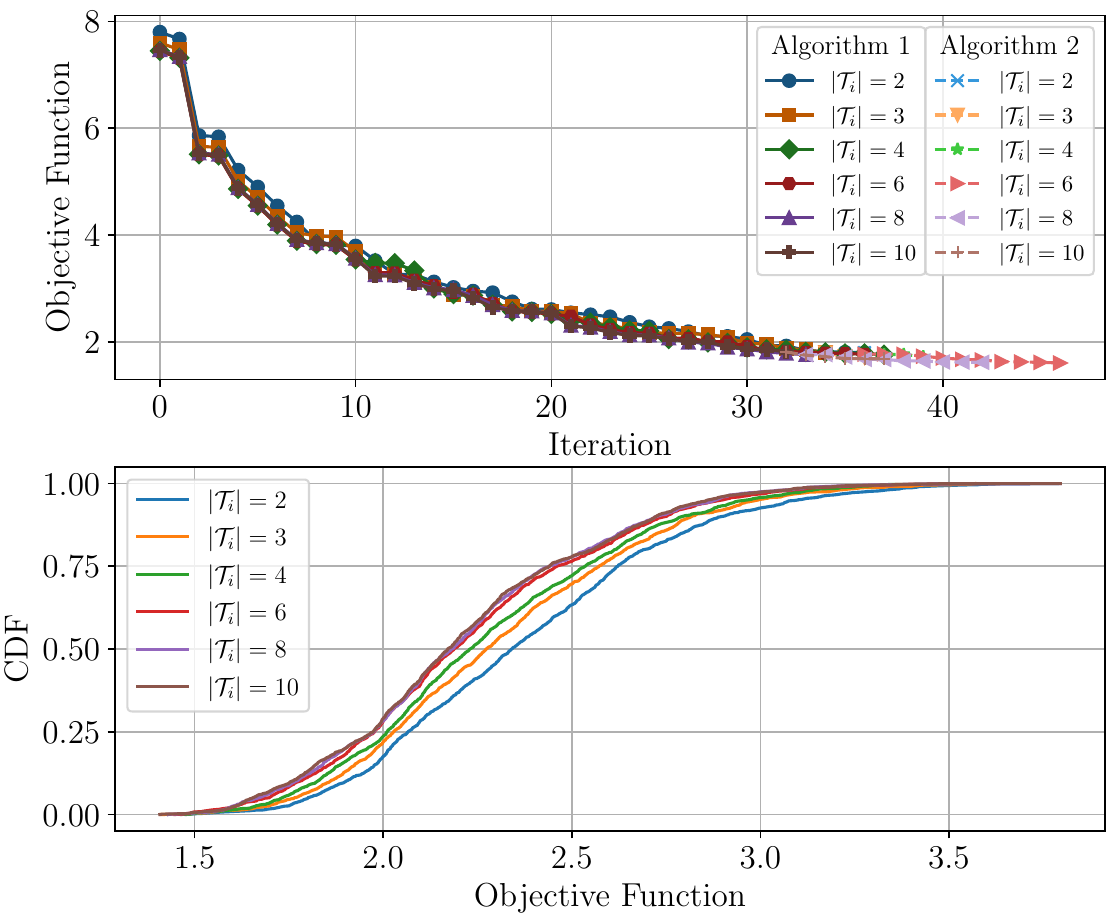}
    \caption{Top plot: Convergence of the objective function with the number of \textit{scheduling periods} $|\mathcal{T}_{ i}|$. Bottom plot: CDF of the objective function values over $|\mathcal{I}|=1800$ assignment  periods.}
    \label{fig:scheduling periods}
\end{figure}

We also analyze the convergence of the proposed heuristic over $|\mathcal{I}|=1800$ assignment periods, considering $40$ moving \glspl{UE}, $20$ \gls{RIS} devices, and \textit{scheduling periods} $|\mathcal{T}_{ i}| \in \left[2,10\right]$ for $\forall i \in \mathcal{I}$. The results, depicted in Fig.~\ref{fig:scheduling periods}, reveal that as the number of \textit{scheduling periods} $|\mathcal{T}_{ i}|$ increases, more iterations are needed to extract a solution in a single assignment period while improving the objective function value. The bottom plot displays the \gls{CDF} of the objective function values collected over the $|\mathcal{I}|=1800$ assignment periods. While increasing the number of \textit{scheduling periods} generally reduces the objective function value, similar performance is achieved for $|\mathcal{T}_{ i}| \geq 6$. Based on these observations, we empirically set $|\mathcal{T}_{ i}| = 10$ \textit{scheduling periods} for the remaining experiments. These results indicate that further subdivision of the assignment period into smaller scheduling periods provides negligible delay-bound improvement, supporting the design choice of keeping each \gls{RIS} configuration fixed within a scheduling period as finer reconfiguration would only yield marginal objective-function improvements.

\begin{figure}[t!]
    \centering
    \includegraphics[width=\columnwidth]{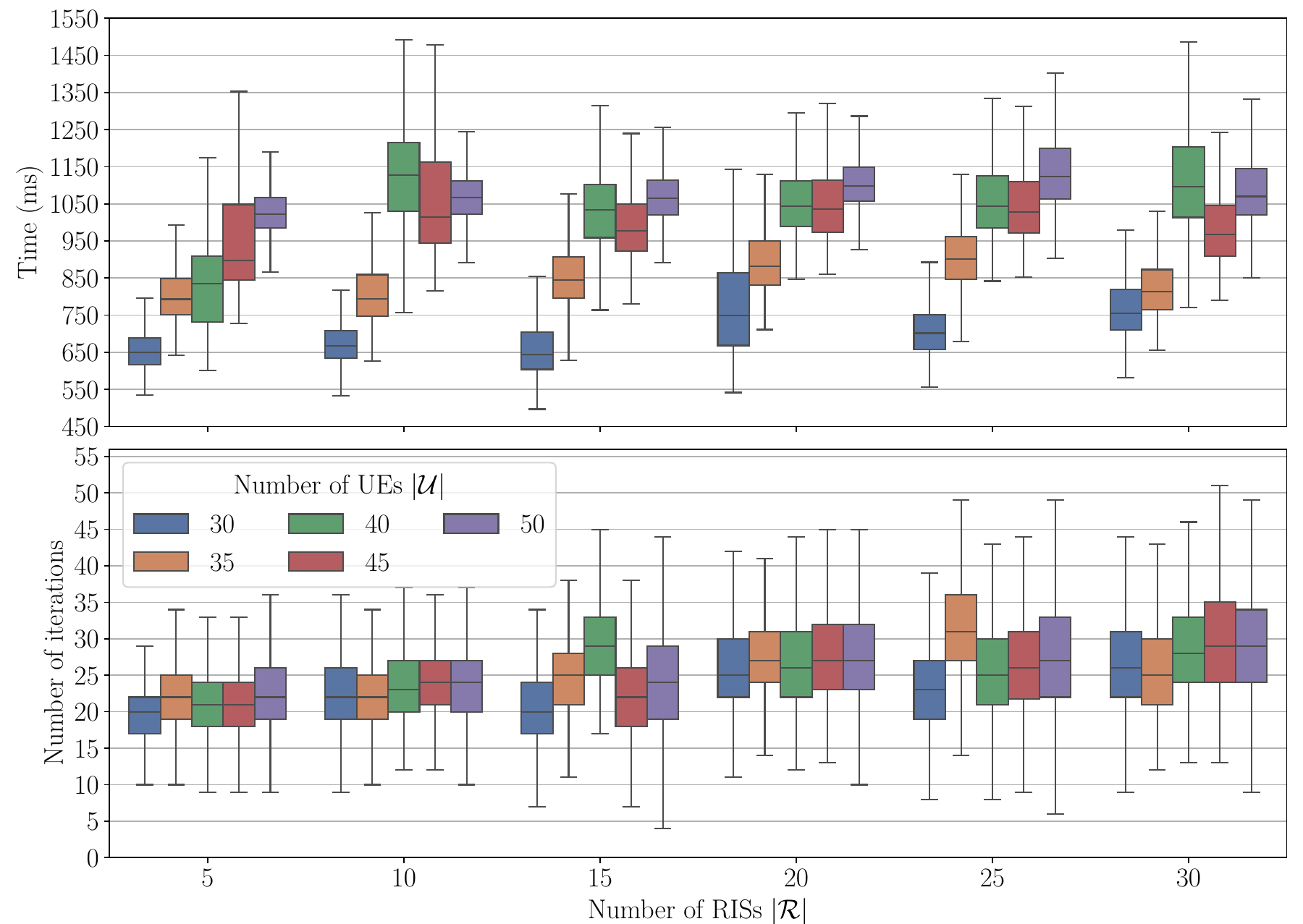}
    \caption{Boxplot representation of the distribution of the execution time of \name{} for different network scenarios.}
    \label{fig:ExecutionTime}
\end{figure}

\subsection{Computational Complexity Analysis of DARIO}
\label{subsec:complexity}

We evaluate the computational complexity of \name{} and the number of iterations required by \name{} to determine the \gls{UE}-\gls{RIS} assignment for an assignment period $i\in \mathcal{I}$. The boxplot in Fig.~\ref{fig:ExecutionTime} shows the distribution of execution time and iteration count across heterogeneous deployment scenarios. The execution time slightly increases with the number of \gls{RIS} devices and \glspl{UE}. This is due to a higher probability of multiple \glspl{UE} having \gls{LOS} with the same group of \gls{RIS} devices, which requires \name{} to explore more candidate solutions to minimize the objective function. The number of iterations typically ranges from $17$ to $35$, with a maximum of $51$. \name{} shows an affordable execution time ($<1.55$~s), below the duration of the considered \textit{assignment period} ($I_{\scriptstyle time}=2$~s). This result confirms that \name{} completes its optimization within the Non-\gls{RT} control interval defined by O-RAN (i.e., $\leq I_{\scriptstyle time}=2$). The remaining $\sim$0.45~s per $I_{\scriptstyle time}$ can be used to disseminate the orchestration outputs, namely the radio resource allocation policies and \gls{RIS} configuration plans, through the proposed \texttt{RO2} and \texttt{RO3} interfaces before the next assignment period begins.

\section{Real-World Evaluation}
\label{sec:real}

We evaluate \name{}'s \emph{rApps}, implemented in Python, using empirical traffic traces collected from a real-world deployment, thereby extending the analysis beyond synthetic simulations. We begin by describing the collected traffic traces, detailing their characteristics and how they were integrated into our evaluation framework to assess \name{}'s performance. Additionally, we outline the methodology used to adapt these traces to align with the network configuration and delay-sensitive requirements considered in this study. Subsequently, we present the  results obtained using these empirical traces, comprehensively comparing \name{} against three reference solutions. This additional testing highlights \name{}'s ability to maintain low-latency guarantees and stable performance under realistic traffic fluctuations.

\subsection{RIS Device: Configuration and Deployment}
\label{subsec:ris}

For the empirical evaluation, we use five commercially available off-the-shelf \gls{RIS} devices~\cite{NEC-ris}. Each \gls{RIS} operates at $5$~GHz and is equipped with a $10 \times 10$ patch antenna array supporting electronic phase-shift control for dynamic beam steering. The devices are managed through a custom Python-based control system developed for this study, which applies \gls{RIS}-wide phase configurations selected from a codebook of $589$ predefined phase profiles, each corresponding to a distinct angular response.

A centralized controller orchestrates time-synchronized configuration across all \gls{RIS} units, ensuring coherent operation during the experiments. Each \gls{RIS} measures $40 \times 40~\mathrm{cm}^2$, which introduces practical deployment constraints related to space availability and \gls{LOS} conditions. To address this, a site-specific assessment was conducted considering coverage, visibility, and environmental reflections, and the \gls{RIS} devices were strategically positioned to maximize beamforming effectiveness.

After deployment, extensive measurements were collected under multiple \gls{RIS} configurations, capturing the resulting signal propagation characteristics. These measurements, used in the subsequent evaluation, provide quantitative evidence of the impact of \gls{RIS}-assisted operation on latency performance and validate \name{} under realistic conditions.

\subsection{Capturing Urban Real-Traces}
\label{subsec:traces}

We collected a dataset from an operational mobile network deployment in an urban area of Madrid, Spain, using advanced traffic tracking tools~\cite{falcon}. From this original dataset, we derived the preprocessed traffic traces used in the real-world evaluation presented in this paper. The released dataset\footnote{The dataset used in this study is publicly available at IEEE DataPort (\url{https://ieee-dataport.org/documents/preprocessed-urban-mobile-traffic-traces}) to ensure reproducibility and facilitate further research.} contains the processed per-user uplink traffic time series at \gls{TTI} granularity employed in our experiments.

To extend the dataset's scope, we conducted measurements under both \gls{RIS}-inactive and \gls{RIS}-active conditions in the network, allowing us to isolate and analyze the impact of RIS configurations on network performance. Specifically, the dataset includes scenarios where RIS devices were inactive, serving as a baseline, and scenarios where RIS devices were dynamically configured to enhance signal propagation. This dual set of measurements enables a detailed comparison, quantifying the performance gains introduced by \gls{RIS}-assisted propagation. To explore large-scale deployments, scenarios with more than five \gls{RIS} devices were emulated using measurement-driven models derived from the physical setup. This ensure the extended experiments remained grounded in realistic, empirically validated configurations and performance metrics.

\begin{figure}[t!]
    \centering
    \includegraphics[width=\columnwidth]{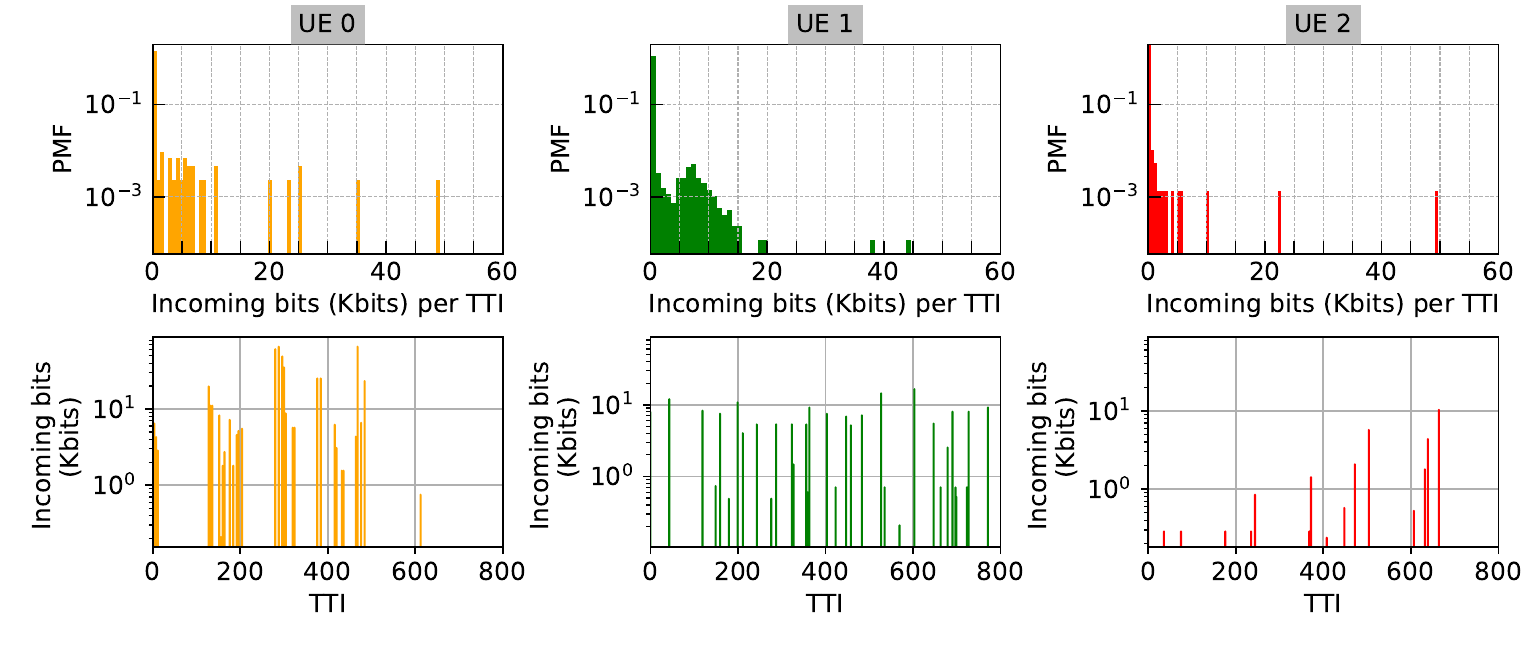}
    \caption{Realization of real traffic traces for 3 UEs. Top: PMF generated traffic. Bottom: Generated traffic per TTI.}
    \label{fig:realtraffic_realization}
\end{figure}

For the experiments, we extracted the \gls{UL} traffic patterns from all \glspl{UE} and estimated the empirical \gls{PMF} of their per-\gls{TTI} packet arrivals. Using these distributions, we randomly selected a subset of \glspl{UE} and synthetically generated traffic following the observed patterns. Fig.~\ref{fig:realtraffic_realization} presents the empirical \glspl{PMF} of three representative \glspl{UE}, along with corresponding synthetic realizations.

By incorporating both baseline and \gls{RIS}-active scenarios, this dataset captures not only empirical mobile network dynamics, but also provides critical insights into the tangible benefits of \gls{RIS} deployment, further supporting the practical applicability of \name{} in urban-based dynamic network environments. 

\subsection{SNC Model Validation Against Real Traces}

This experiment validates the proposed \gls{SNC} delay model using the empirical traffic traces collected from the urban deployment previously described. In this evaluation, the collected traces are used as input to a controlled emulation framework that reproduces the observed traffic dynamics while computing the resulting end-to-end delays under the modeled 5G-\gls{RIS} configuration. The goal is to verify that the model provides reliable and consistent delay-bound estimations under realistic, trace-driven conditions and to demonstrate that its performance remains stable when compared to a synthetic scenario driven by Poisson arrivals.  
\begin{figure}[t!]
    \centering
    \includegraphics[width=\columnwidth]{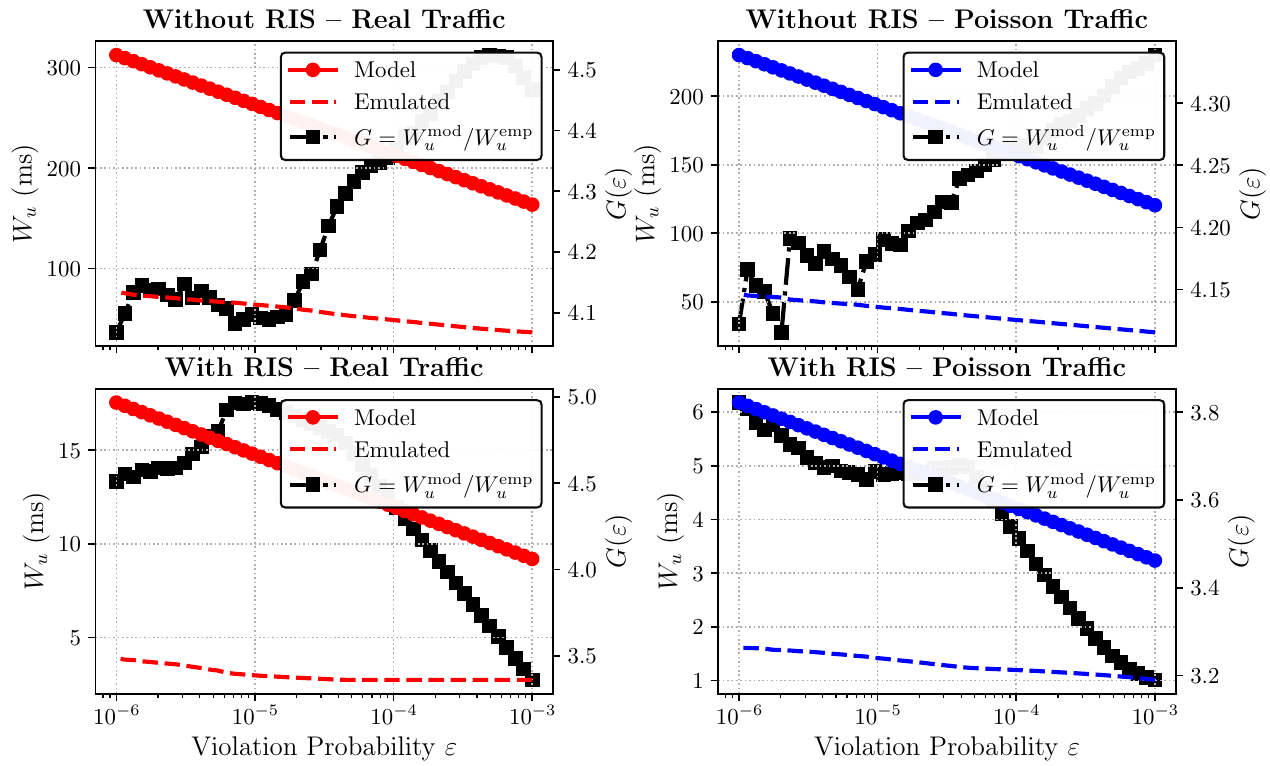}
    \caption{Analytical and emulated delay bounds $W_u^{\mathrm{mod}}$ and $W_u^{\mathrm{emp}}$ as functions of the violation probability $\varepsilon$. The secondary axis depicts the ratio $G(\varepsilon) = W_u^{\mathrm{mod}} / W_u^{\mathrm{emp}}$.}
    \label{fig:realtraffic_validation}
\end{figure}

\begin{figure*}[t!]
    \centering
    \includegraphics[width=\textwidth]{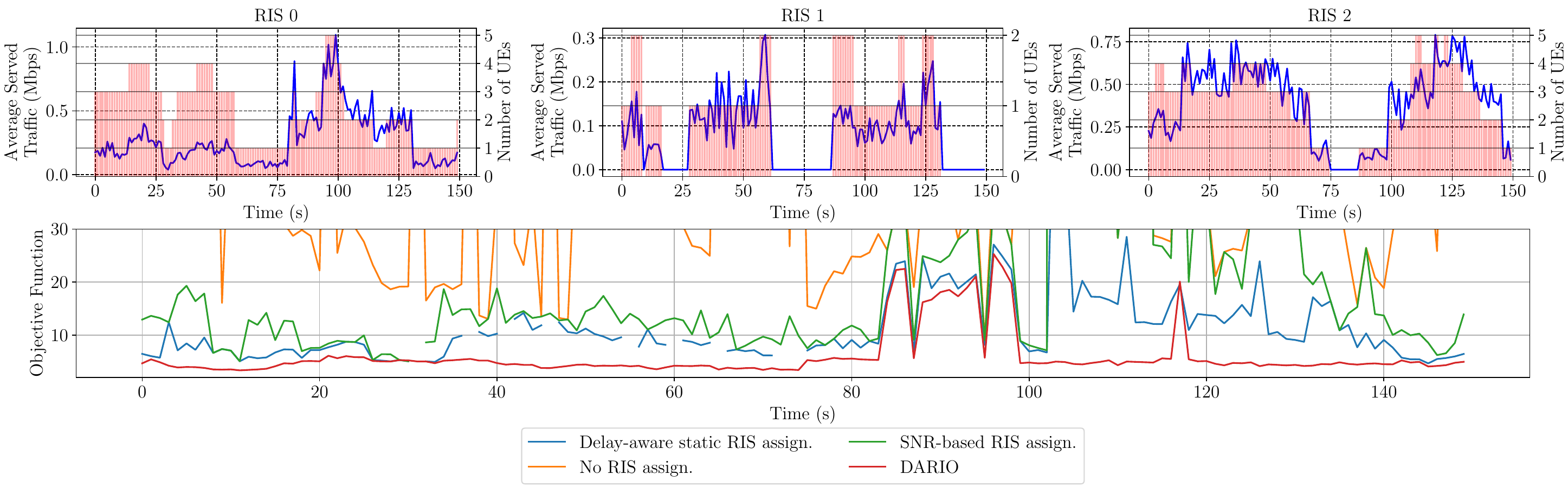}
    \caption{Top plot: Evolution of the average traffic load potentially served by three arbitrary RIS devices. Bottom plot: Evolution of the objective function over the first 160 assignment periods. For these measurements, we consider a scenario with $|\mathcal{U}| = 40$ \glspl{UE} and $|\mathcal{R}| = 15$ \gls{RIS} devices.}
    \label{fig:EvolutionRISassignment_realtraffic}
\end{figure*}

For both traffic types, the analytical delay bound $W_u^{\mathrm{mod}}(\varepsilon)$ and the emulated delay $W_u^{\mathrm{emu}}(\varepsilon)$ were evaluated as functions of the violation probability $\varepsilon \in [10^{-6},10^{-3}]$, considering a single 5G \gls{BS} with and without \gls{RIS} assistance. The comparison also includes the ratio $G(\varepsilon)=W_u^{\mathrm{mod}}(\varepsilon)/W_u^{\mathrm{emu}}(\varepsilon)$, which quantifies the conservatism of the analytical bound with respect to the delay obtained through trace-driven emulation.  

Figure~\ref{fig:realtraffic_validation} shows the obtained results. For both real-trace and Poisson traffic, the analytical and emulated delay curves exhibit a similar evolution across all values of $\varepsilon$, with $G(\varepsilon)$ consistently taking values between 3 and 5. This behaviour indicates that the delay-bound estimation produced by the \gls{SNC} model remains consistent regardless of whether the input follows a memoryless Poisson process or a more complex real-trace pattern. In particular, the model, whose arrival envelope is derived from the empirical \gls{MGF}, responds similarly under both traffic types, suggesting that its operation is largely insensitive to the specific statistical structure of the incoming traffic, including burstiness and temporal correlations.

A slight increase of $G(\varepsilon)$ at low violation probabilities is observed in the \gls{RIS}-assisted case, which is explained by the higher variability of the service process resulting from the cascaded \gls{UE}–\gls{RIS}–\gls{BS} channel and the quantization-induced phase errors. This variability affects the tail behavior of the delay distribution, slightly widening the analytical bound.

\subsection{Performance Evaluation of \name{} against Reference Solutions}
\label{subsec:testing}
We evaluate the performance of \name{} using the real traffic traces described in Section~\ref{subsec:traces}, thereby assessing its practical effectiveness under realistic and time-varying network conditions. Since there are no existing \gls{RIS} control frameworks that jointly address dynamic \gls{UE}-\gls{RIS} assignment, user mobility, stochastic traffic arrivals, and probabilistic delay-bound guarantees, we compare \name{} against three benchmark solutions: $i$) No RIS Assignment: The baseline scenario where the \gls{BS} operates without any \gls{UE}-\gls{RIS} assignment. This configuration isolates the performance gains attributable solely to the integration of \gls{RIS} devices;
$ii$) SNR-Based RIS Assignment: A static, one-to-one \gls{UE}-\gls{RIS} assignment strategy, where each \gls{UE} is paired with the \gls{RIS} providing the highest average \gls{SNR} over the \gls{UE}-\gls{RIS}-\gls{BS} link. The static nature of this approach means the assignment remains fixed throughout all \textit{scheduling periods} $t \in \mathcal{T}_{ i}$ within a single \textit{assignment period} $i \in \mathcal{I}$; $iii$) Delay-Aware Static RIS Assignment: A partially delay-aware solution that considers \gls{UE} delay requirements for one-to-one \gls{UE}-\gls{RIS} assignment during the entire \textit{assignment period} $i$. However, it lacks the flexibility of \name{} in supporting dynamic, many-to-many \gls{UE}-\gls{RIS} assignments across \textit{scheduling periods}. All these schemes are evaluated under identical channel and CSI assumptions, so the observed performance differences are attributable to the orchestration and scheduling strategies.

To evaluate \name{}, we measured the objective function value over $|\mathcal{I}|=1800$ \textit{assignment periods} under varying configurations of \glspl{UE} and \glspl{RIS}. Fig.~\ref{fig:EvolutionRISassignment_realtraffic} illustrates network dynamics during the first 160 assignment periods for a scenario with $|\mathcal{U}|=40$ \glspl{UE} and $|\mathcal{R}|=15$ \glspl{RIS}. The top plot shows average traffic (blue curve) served by three RIS devices, alongside the number of \glspl{UE} (red bars) with \gls{LOS} to those RIS devices over time, highlighting the dynamic traffic conditions \name{} is designed to handle. The bottom plot shows the temporal evolution of the objective function, where \gls{RIS}-enabled solutions (green, blue, and red curves) consistently outperform the baseline (orange curve) across all assignment periods. Notably, \name{} (red curve) yields the lowest objective function values throughout, highlighting its superior adaptability to traffic dynamics. 

\begin{figure}[t!]
    \centering
    \includegraphics[clip, trim=0cm 0cm 0cm 0cm, width=\columnwidth]{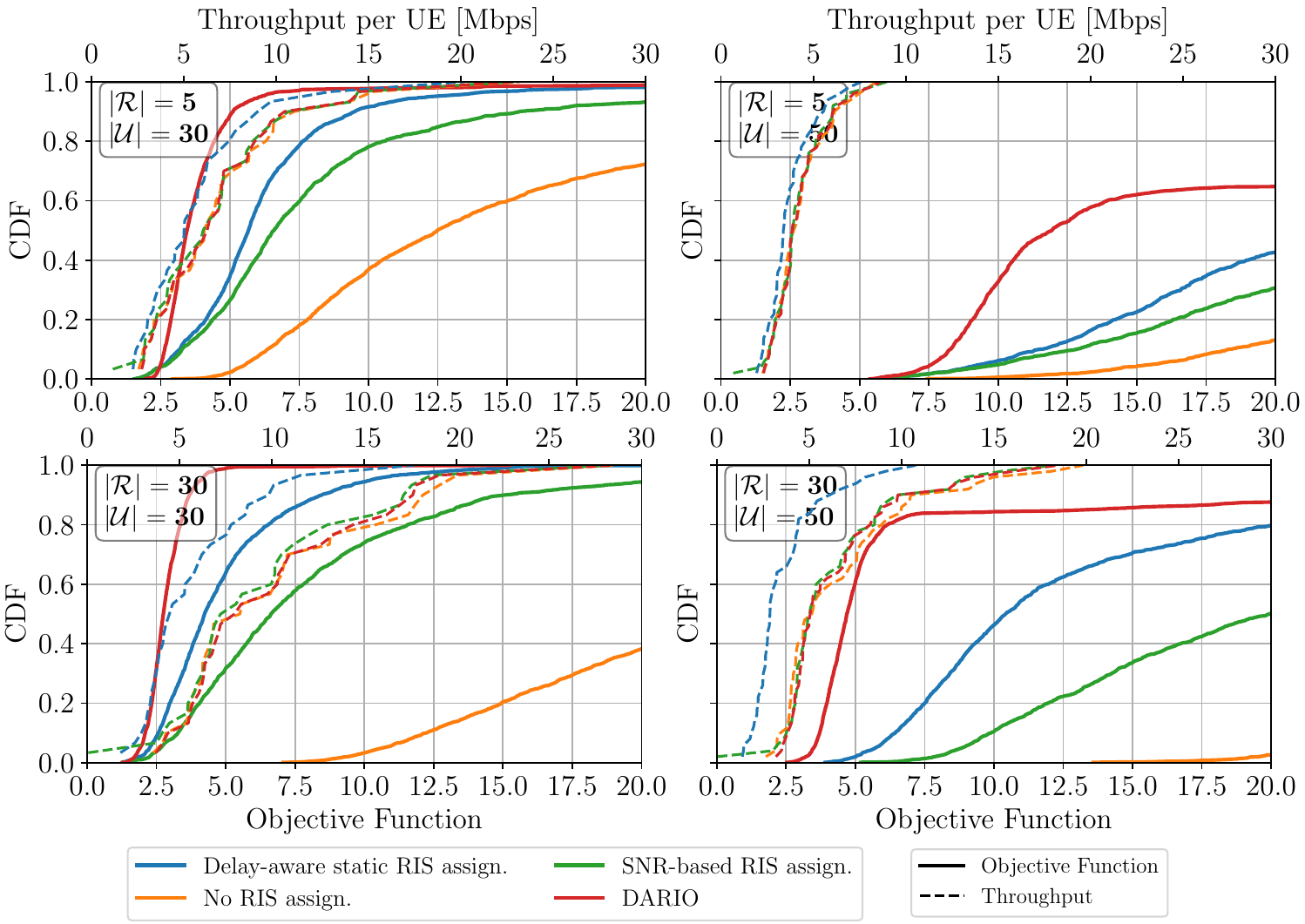}
    \caption{CDFs of the values of the objective function when it is measured over $|\mathcal{I}|=1800$ assignment periods for different network scenarios.}
    \label{fig:EvolutionRISassignmentCDF_realtraffic}
\end{figure}

To provide a holistic comparison, we also evaluated \name{} across different scenarios where \glspl{UE} and \glspl{RIS} vary in number. Fig.~\ref{fig:EvolutionRISassignmentCDF_realtraffic} presents the \glspl{CDF} of the objective function under four extreme scenarios: Scenario ($i$): $|\mathcal{U}| = 30$, $|\mathcal{R}| = 5$ (few \glspl{UE} and RIS); Scenario ($ii$): $|\mathcal{U}| = 50$, $|\mathcal{R}| = 5$ (many \glspl{UE} and few RIS); Scenario ($iii$): $|\mathcal{U}| = 30$, $|\mathcal{R}| = 30$ (few \glspl{UE} and many RIS); and Scenario ($iv$): $|\mathcal{U}| = 50$, $|\mathcal{R}| = 30$ (many \glspl{UE} and many RIS).

In all scenarios, \name{} consistently outperforms the other strategies, followed by Delay-Aware Static, SNR-Based, and finally No RIS. This ordering is expected: No RIS does not leverage RIS technology; SNR-Based, although using RIS, ignores delay requirements; and among the delay-aware approaches, \name{} benefits from dynamic many-to-many \gls{UE}--\gls{RIS} assignment within each \textit{assignment period}, unlike Delay-Aware Static, which uses fixed mappings.  

In Scenario~(\textit{i}), all \glspl{CDF} are closely aligned due to the low number of \glspl{UE} and \gls{RIS} devices, which limits the likelihood of a \gls{UE} experiencing \gls{LOS} with a \gls{RIS} during an \textit{assignment period}. Consequently, RIS-enabled strategies provide only modest gains over No RIS. Even so, \name{} achieves a 90th percentile (P90) gain of 79.08\% over No RIS, 68.99\% over SNR-Based, and 52.20\% over Delay-Aware Static; 50th percentile (P50) gains are 64.90\%, 46.30\%, and 36.62\%, respectively.  

In Scenario~(\textit{ii}), the number of \glspl{UE} increases while \gls{RIS} count remains unchanged, raising network load and shifting all \glspl{CDF} to the right. Here, \name{}’s P90 gains are smaller than in Scenario~(\textit{i}), 5.61\% over No RIS, 47.66\% over SNR-Based, and 16.53\% over Delay-Aware Static, because more \glspl{UE} compete for the same \gls{RIS} devices. Nonetheless, P50 gains (81.73\%, 68.24\%, and 47.37\%, respectively) still show that \name{} provides a clear advantage.  

In Scenario~(\textit{iii}), \gls{RIS} availability increases substantially, shifting the \glspl{CDF} of RIS-enabled solutions to the left, while No RIS moves further right. With a nearly 1:1 \gls{UE}--\gls{RIS} ratio, all RIS-based approaches benefit, but the dynamic mapping of \name{} extracts more value from a better use of \gls{RIS} devices than static methods. As a result, \name{}’s P90 gains (95.71\%, 73.32\%, and 55.10\% over No RIS, SNR-Based, and Delay-Aware Static, respectively) are larger than in Scenarios~(\textit{i}) and (\textit{ii}). P50 gains are 87.76\%, 55.52\%, and 32.55\%, respectively.  

Finally, Scenario~(\textit{iv}) consider the higher number of \glspl{UE} and \gls{RIS} devices, increasing load relative to Scenario~(\textit{iii}) but also creating more opportunities for dynamic \gls{UE}-\gls{RIS} mapping. Here, the performance gaps between RIS-based strategies widen, with \name{} achieving substantial P90 gains over No RIS (78.38\%), SNR-Based (75.82\%), and Delay-Aware Static (46.61\%). P50 gains reach 91.34\%, 72.40\%, and 52.35\%, respectively, confirming that per-\textit{scheduling-period} \gls{RIS} reassignment is particularly beneficial under heavy load.

In addition, Fig.~\ref{fig:EvolutionRISassignmentCDF_realtraffic} also reports the \gls{CDF} of the average per-\gls{UE} throughput (dashed curves, top x-axis), where curves shifted to the right indicate higher throughput. Across all scenarios, \name{} achieves per-\gls{UE} throughput comparable to the SNR-based and no-\gls{RIS} baselines, while consistently outperforming Delay-Aware Static \gls{RIS} assignment at medium and high percentiles. This confirms that the delay-bound improvements provided by \name{} are not obtained at the expense of a significant throughput loss: dynamic \gls{RIS} reallocation primarily reshapes the service process to reduce delay violations, while preserving the long-term average throughput experienced by the \glspl{UE}.

\section{Conclusions}
\label{sec:Conclusions}

In this paper, we presented \name{}, an \gls{O-RAN}-compliant \gls{RIS} scheduler that dynamically configures multiple \gls{RIS} devices across time slots to establish controlled \gls{LOS} links between \glspl{UE} and the \gls{BS}. By adaptively assigning \glspl{UE} to \gls{RIS} devices, \name{} reduces packet delays and supports delay-sensitive requirements under dynamic channel and traffic conditions. This is enabled by a novel \gls{SNC}-based model for estimating delay violation probabilities, which guides near-optimal \gls{UE}--\gls{RIS} assignments formulated as a \gls{NIP} and solved via a computationally efficient heuristic.

We validated \name{} using real traffic traces and measurements from an operational deployment with five commercial \gls{RIS} devices, and leveraged this data to evaluate larger-scale scenarios. The results demonstrate the scalability and practical benefits of \name{} in realistic environments, consistently outperforming all benchmark strategies across all scenarios. Performance gains depend on traffic load and \gls{RIS} availability, reaching up to 95.71\% over No RIS, 75.82\% over SNR-Based, and 55.10\% over Delay-Aware Static at the 90th percentile, thereby confirming the advantage of dynamic many-to-many \gls{UE}--\gls{RIS} assignment over static approaches.

\bibliographystyle{IEEEtran}
\bibliography{references}

@ARTICLE{Fidler1,
  author={M. {Fidler} and A. {Rizk}},
  journal={IEEE Commun. Surveys Tuts.}, 
  title={{A} {G}uide to the {S}tochastic {N}etwork {C}alculus}, 
  year={2015},
  volume={17},
  number={1},
  pages={92-105},
  doi={10.1109/COMST.2014.2337060}}

@book{ross2014first,
  title={{A} {F}irst {C}ourse in {P}robability},
  author={Ross, Sheldon},
  year={2014},
  publisher={Pearson}
}

@ARTICLE{Kishk2021,
  author={Kishk, Mustafa A. and Alouini, Mohamed-Slim},
  journal={IEEE J. Sel. Areas Commun.}, 
  title={{Exploiting Randomly Located Blockages for Large-Scale Deployment of Intelligent Surfaces}}, 
  year={2021},
  volume={39},
  number={4},
  pages={1043-1056},
  doi={10.1109/JSAC.2020.3018808}}

@ARTICLE{Du2021,
  author={Du, Linsong and Shao, Shihai and Yang, Gang and Ma, Jianhui and Liang, Qingpeng and Tang, Youxi},
  journal={IEEE Trans. Wireless Commun.}, 
  title={{Capacity Characterization for Reconfigurable Intelligent Surfaces Assisted Multiple-Antenna Multicast}}, 
  year={2021},
  volume={20},
  number={10},
  pages={6940-6953},
  doi={10.1109/TWC.2021.3078853}}

@ARTICLE{Guo2020,
  author={Guo, Huayan and Liang, Ying-Chang and Chen, Jie and Larsson, Erik G.},
  journal={IEEE Trans. Wireless Commun.}, 
  title={{Weighted Sum-Rate Maximization for Reconfigurable Intelligent Surface Aided Wireless Networks}}, 
  year={2020},
  volume={19},
  number={5},
  pages={3064-3076},
  doi={10.1109/TWC.2020.2970061}}

@ARTICLE{Du2021-2,
  author={Du, Linsong and Zhang, Wei and Ma, Jianhui and Tang, Youxi},
  journal={IEEE Trans. Veh. Technol.}, 
  title={{Reconfigurable Intelligent Surfaces for Energy Efficiency in Multicast Transmissions}}, 
  year={2021},
  volume={70},
  number={6},
  pages={6266-6271},
  doi={10.1109/TVT.2021.3080302}}

@ARTICLE{Xue2024,
  author={Xue, Qing and Xia, Hao and Mu, Jiajun and Xu, Yongjun and Yan, Li and Ma, Shaodan},
  journal={IEEE Trans. Green Commun. Netw.}, 
  title={{User-Centric Association for Dense mmWave Communication Systems With Multi-Connectivity}}, 
  year={2024},
  volume={8},
  number={1},
  pages={177-189},
  doi={10.1109/TGCN.2023.3332773}}

@ARTICLE{Liu2023,
  author={Liu, Sifan and Liu, Rang and Li, Ming and Liu, Yang and Liu, Qian},
  journal={IEEE Trans. Veh. Technol.}, 
  title={{Joint BS-RIS-User Association and Beamforming Design for RIS-Assisted Cellular Networks}}, 
  year={2023},
  volume={72},
  number={5},
  pages={6113-6128},
  doi={10.1109/TVT.2022.3231347}}

@INPROCEEDINGS{Manhattanmodel,
  author={Bai, F. and Narayanan Sadagopan and Helmy, A.},
  booktitle={IEEE INFOCOM}, 
  title={{IMPORTANT: a framework to systematically analyze the Impact of Mobility on Performance of Routing Protocols for Adhoc Networks}}, 
  year={2003},
  volume={2},
  number={},
  pages={825-835 vol.2},
  doi={10.1109/INFCOM.2003.1208920}}

@ARTICLE{Saad2020,
  author={Saad, Walid and Bennis, Mehdi and Chen, Mingzhe},
  journal={IEEE Netw.}, 
  title={{A Vision of 6G Wireless Systems: Applications, Trends, Technologies, and Open Research Problems}}, 
  year={2020},
  volume={34},
  number={3},
  pages={134-142},
  doi={10.1109/MNET.001.1900287}}

@Inbook{Hemmecke2010,
author="Hemmecke, Raymond
and K{\"o}ppe, Matthias
and Lee, Jon
and Weismantel, Robert",
title="{Nonlinear Integer Programming}",
bookTitle="50 Years of Integer Programming 1958-2008: From the Early Years to the State-of-the-Art",
year="2010",
publisher="Springer Berlin Heidelberg",
address="Berlin, Heidelberg",
pages="561--618",
isbn="978-3-540-68279-0",
doi="10.1007/978-3-540-68279-0_15",
url="https://doi.org/10.1007/978-3-540-68279-0_15"
}

@INPROCEEDINGS{Mukherjee2022,
  author={Mukherjee, Mithun and Kumar, Vikas and Kumar, Suman and Mavromoustakis, C. X. and Zhang, Qi and Guo, Mian},
  booktitle={IEEE Globecom}, 
  title={{RIS-assisted Task Offloading for Wireless Dead Zone to Minimize Delay in Edge Computing}}, 
  year={2022},
  volume={},
  number={},
  pages={2554-2559},
  doi={10.1109/GLOBECOM48099.2022.10001478}}

@INPROCEEDINGS{Xia2022,
  author={Xia, Lu and others},
  booktitle={IEEE Globecom}, 
  title={{Delay Minimization for RIS-NOMA Assisted MEC Networks With SWIPT}}, 
  year={2022},
  volume={},
  number={},
  pages={4631-4636},
  doi={10.1109/GLOBECOM48099.2022.10001448}}

@ARTICLE{Almekhlafi2022,
  author={Almekhlafi, Mohammed and Arfaoui, Mohamed Amine and Elhattab, Mohamed and Assi, Chadi and Ghrayeb, Ali},
  journal={IEEE Trans. Commun.}, 
  title={{Joint Resource Allocation and Phase Shift Optimization for RIS-Aided eMBB/URLLC Traffic Multiplexing}}, 
  year={2022},
  volume={70},
  number={2},
  pages={1304-1319},
  doi={10.1109/TCOMM.2021.3127265}}

@ARTICLE{Polese2023,
  author={Polese, Michele and Bonati, Leonardo and D’Oro, Salvatore and Basagni, Stefano and Melodia, Tommaso},
  journal={IEEE Commun. Surv. Tutor.}, 
  title={Understanding O-RAN: Architecture, Interfaces, Algorithms, Security, and Research Challenges}, 
  year={2023},
  volume={25},
  number={2},
  pages={1376-1411},
  doi={10.1109/COMST.2023.3239220}}

@ARTICLE{Anand2020,
  author={Anand, Arjun and de Veciana, Gustavo and Shakkottai, Sanjay},
  journal={IEEE/ACM Trans. Netw.}, 
  title={{Joint Scheduling of URLLC and eMBB Traffic in 5G Wireless Networks}}, 
  year={2020},
  volume={28},
  number={2},
  pages={477-490},
  doi={10.1109/TNET.2020.2968373}}

@INPROCEEDINGS{Jiang2023,
  author={Jiang, Wei and Schotten, Hans D.},
  booktitle={IEEE Globecom}, 
  title={{A Simple Multiple-Access Design for Reconfigurable Intelligent Surface-Aided Systems}}, 
  year={2023},
  volume={},
  number={},
  pages={2414-2419},
  doi={10.1109/GLOBECOM54140.2023.10437343}}

@ARTICLE{Abu-Ali2014,
  author={Abu-Ali, Najah and Taha, Abd-Elhamid M. and Salah, Mohamed and Hassanein, Hossam},
  journal={IEEE Commun. Surv. Tutor.}, 
  title={{Uplink Scheduling in LTE and LTE-Advanced: Tutorial, Survey and Evaluation Framework}}, 
  year={2014},
  volume={16},
  number={3},
  pages={1239-1265},
  doi={10.1109/SURV.2013.1127.00161}}

@ARTICLE{Rossanese2024,
  author={Rossanese, Marco and Mursia, Placido and Garcia-Saavedra, Andres and Sciancalepore, Vincenzo and Asadi, Arash and Costa-Perez, Xavier},
  journal={ IEEE Internet Comput.}, 
  title={{Open Experimental Measurements of Sub-6GHz Reconfigurable Intelligent Surfaces}}, 
  year={2024},
  volume={28},
  number={2},
  pages={19-28},
  doi={10.1109/MIC.2024.3376772}}

@inproceedings{Adamuz2024,
  author    = {O. Adamuz-Hinojosa and L. Zanzi and V. Sciancalepore and A. Garcia-Saavedra and X. Costa-Pérez},
  title     = {{ORANUS: Latency-tailored Orchestration via Stochastic Network Calculus in 6G O-RAN}},
  booktitle = {IEEE INFOCOM},
  year      = {2024}
}

@ARTICLE{He2018,
  author={He, Jianhua and Tang, Zuoyin and Ding, Zhiguo and Wu, Dapeng},
  journal={IEEE Trans. Veh. Technol.}, 
  title={{Successive Interference Cancellation and Fractional Frequency Reuse for LTE Uplink Communications}}, 
  year={2018},
  volume={67},
  number={11},
  pages={10528-10542},
  doi={10.1109/TVT.2018.2865814}}

@ARTICLE{Souza2024,
  author={De Souza, Joao Henrique Inacio and Croisfelt, Victor and Kotaba, Radosław and Abrão, Taufik and Popovski, Petar},
  journal={IEEE Commun. Lett.}, 
  title={{Uplink Multiplexing of eMBB/URLLC Services Assisted by Reconfigurable Intelligent Surfaces}}, 
  year={2024},
  volume={},
  number={},
  pages={1-1},
  doi={10.1109/LCOMM.2024.3426613}}

@book{jiang2008stochastic,
  title={{Stochastic Network Calculus}},
  author={Jiang, Yuming and others},
  volume={1},
  year={2008},
  publisher={Springer},
  address={London, U.K.}
}

@book{leboudec2001network,
  title={{Network Calculus: A Theory of Deterministic Queuing Systems for the Internet}},
  author={Le Boudec, Jean-Yves and Thiran, Patrick},
  volume={2050},
  year={2001},
  publisher={Springer},
  address={Berlin, Germany}
}

@techreport{3gpp-ts-38214,
  title        = {{NR; Physical layer procedures for data (Release 18)}},
  author       = {{3GPP}},
  year         = {2024},
  month        = {June},
  number       = {V18.3.0},
  type         = {Technical Specification (TS)}
}

@ARTICLE{Fidler2010,
  author={Fidler, Markus},
  journal={IEEE Commun. Surv. Tutor.}, 
  title={{Survey of deterministic and stochastic service curve models in the network calculus}}, 
  year={2010},
  volume={12},
  number={1},
  pages={59-86},
  doi={10.1109/SURV.2010.020110.00019}}

@techreport{3gpp-ts-38901,
  title        = {{Study on channel model for frequencies from 0.5 to 100 GHz (Release 18)}},
  author       = {{3GPP}},
  year         = {2024},
  month        = {March},
  number       = {V18.0.0},
  type         = {Technical Report (TR)}
}

@INPROCEEDINGS{falcon,
  author={Falkenberg, Robert and Wietfeld, Christian},
  booktitle={2019 IEEE GLOBECOM}, 
  title={{FALCON}: {A}n {A}ccurate {R}eal-{T}ime {M}onitor for {C}lient-{B}ased {M}obile {N}etwork {D}ata {A}nalytics}, 
  year={2019},
  volume={},
  number={},
  pages={1-7},
  doi={10.1109/GLOBECOM38437.2019.9014096}}

@misc{NEC-ris,
    author = {{NEC Corporation}},
    title = {{NEC Smart Surface v2022}},
    howpublished = {\url{https://www.nec-enterprise.com/it/projects}},
    year = 2023
}

@article{WCL2024_RateRegion,
  title={Rate Region of {RIS}-Aided {URLLC} Broadcast Channels: Diagonal versus Beyond Diagonal Globally Passive {RIS}},
  author={Soleymani, M. and Zappone, A. and Jorswieck, E. and Di Renzo, M. and Santamaria, I.},
  journal={IEEE Wirel. Commun. Lett.},
  doi={10.1109/LWC.2024.3500100},
  year={2024},
  publisher={IEEE}
}

@article{Soleymani2024_RSMA,
  title={Rate Splitting Multiple Access for RIS-aided URLLC MIMO Broadcast Channels},
  author={Soleymani, Mohammad and Santamaria, Ignacio and Jorswieck, Eduard and Clerckx, Bruno},
  journal={arXiv preprint arXiv:2411.11028},
  year={2024}
}

@article{Liu2021_UplinkMultiplexing,
  title={Uplink Multiplexing of eMBB/URLLC Services Assisted by Reconfigurable Intelligent Surfaces},
  author={Liu, Liang and Yu, Wei},
  journal={IEEE J. Sel. Areas Commun.},
  volume={39},
  number={7},
  pages={1931--1945},
  year={2021},
  publisher={IEEE},
  doi={10.1109/JSAC.2021.1234567}
}

@ARTICLE{Peng2024,
  author={Peng, Qihao and Ren, Hong and Pan, Cunhua and Elkashlan, Maged and García Armada, Ana and Popovski, Petar},
  journal={IEEE Trans. Wireless Commun.}, 
  title={{Two-Timescale Design for Reconfigurable Intelligent Surface-Aided URLLC}}, 
  year={2024},
  volume={23},
  number={10},
  pages={13664-13677},
  doi={10.1109/TWC.2024.3403929}}

@ARTICLE{Sang2024,
  author={Sang, Jian and Lan, Jifeng and Zhou, Mingyong and Gao, Boning and Tang, Wankai and Li, Xiao and Matthaiou, Michail and Jin, Shi and Renzo, Marco Di},
  journal={IEEE Transactions on Vehicular Technology}, 
  title={Measurement-Based Small-Scale Channel Model for Sub-6 GHz RIS-Assisted Communications}, 
  year={2024},
  volume={73},
  number={8},
  pages={12178-12183},
  doi={10.1109/TVT.2024.3373819}}

@ARTICLE{Xu2021,
  author={Xu, Peng and Chen, Gaojie and Yang, Zheng and Renzo, Marco Di},
  journal={IEEE Wireless Communications Letters}, 
  title={{Reconfigurable Intelligent Surfaces-Assisted Communications With Discrete Phase Shifts: How Many Quantization Levels Are Required to Achieve Full Diversity?}}, 
  year={2021},
  volume={10},
  number={2},
  pages={358-362},
  doi={10.1109/LWC.2020.3031084}}

@article{Badiu2019,
  author    = {M. A. Badiu and J. P. Coon},
  title     = {{Communication through a large reflecting surface with phase errors}},
  journal   = {IEEE Wireless Communications Letters},
  year      = {2019},
  volume    = {9},
  number    = {2},
  pages     = {184--188},
}

@ARTICLE{Adamuz-Hinojosa-MAREA,
  author={Adamuz-Hinojosa, Oscar and Zanzi, Lanfranco and Sciancalepore, Vincenzo and Costa-Pérez, Xavier},
  journal={IEEE Trans. Commun.}, 
  title={{MAREA: A Delay-Aware Multi-time-Scale Radio Resource Orchestrator for 6G O-RAN}}, 
  year={2025},
  volume={},
  number={},
  pages={1-1},
  doi={10.1109/TCOMM.2025.3552296}}

@ARTICLE{Adamuz-Hinojosa-SNC,
  author={Adamuz-Hinojosa, Oscar and Sciancalepore, Vincenzo and Ameigeiras, Pablo and Lopez-Soler, Juan M. and Costa-Pérez, Xavier},
  journal={IEEE Trans. Wireless Commun.}, 
  title={{A Stochastic Network Calculus (SNC)-Based Model for Planning B5G uRLLC RAN Slices}}, 
  year={2023},
  volume={22},
  number={2},
  pages={1250-1265},
  doi={10.1109/TWC.2022.3203937}}

@ARTICLE{Haochen2024,
  author={Li, Haochen and Zhiwen, Pan and Bin, Wang and Nan, Liu and Xiaohu, You},
  journal={IEEE Internet Things J.}, 
  title={{Channel Estimation for Reconfigurable-Intelligent-Surface-Aided Multiuser Communication Systems Exploiting Statistical CSI of Correlated RIS–User Channels}}, 
  year={2024},
  volume={11},
  number={5},
  pages={8871-8881},
  doi={10.1109/JIOT.2023.3321931}}

@INPROCEEDINGS{Cheng2024,
  author={Cheng, Hai and Johari, Pedram and Arfaoui, Mohamed Amine and Periard, Francois and Pietraski, Philip and Zhang, Guodong and Melodia, Tommaso},
  booktitle={WONS}, 
  title={{Real-Time AI-Enabled CSI Feedback Experimentation with Open RAN}}, 
  year={2024},
  volume={},
  number={},
  pages={121-124},
  doi={10.23919/WONS60642.2024.10449603}}

@ARTICLE{Yao-1,
  author={Yao, Yu and others},
  journal={IEEE Trans. Commun.}, 
  title={{Hybrid RIS-Enhanced ISAC Secure Systems: Joint Optimization in the Presence of an Extended Target}}, 
  year={2025},
  volume={73},
  number={12},
  pages={15688-15704},
  doi={10.1109/TCOMM.2025.3610218}}

@ARTICLE{Yao-2,
  author={Yao, Yu and Zhu, Zhixing and Miao, Pu and Cheng, Xu and Shu, Feng and Wang, Jiangzhou},
  journal={IEEE Trans. Veh. Technol.}, 
  title={{Optimizing Hybrid RIS-Aided ISAC Systems in V2X Networks: A Deep Reinforcement Learning Method for Anti-Eavesdropping Techniques}}, 
  year={2025},
  volume={74},
  number={6},
  pages={9224-9239},
  doi={10.1109/TVT.2025.3538471}}

\vskip -2\baselineskip plus -1fil
\begin{IEEEbiography}
[{\includegraphics[width=1in,height=1.25in,clip,keepaspectratio]{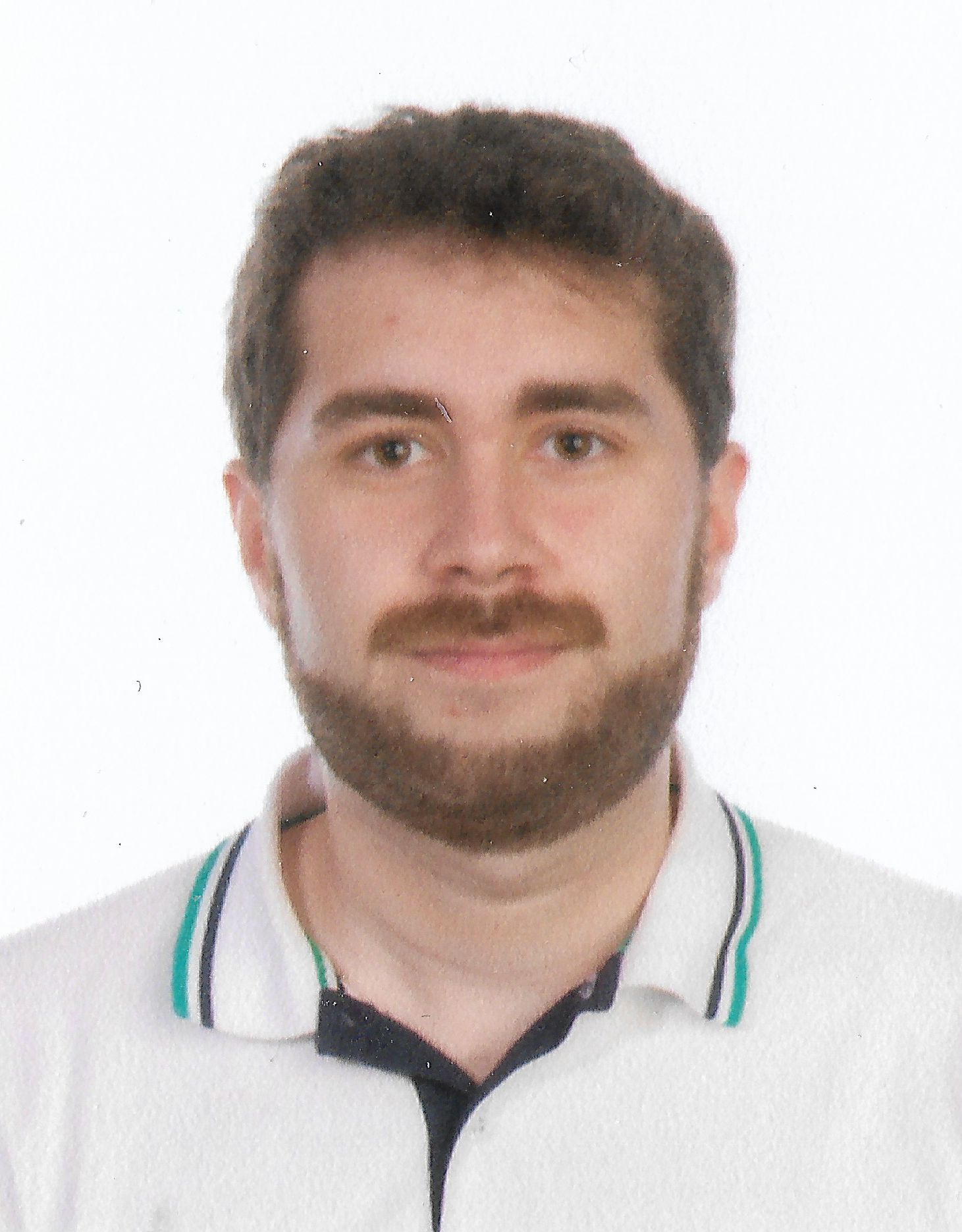}}]
{Oscar Adamuz-Hinojosa } received the B.Sc., M.Sc., and Ph.D. degrees in telecommunications engineering from the University of Granada, Granada, Spain, in 2015, 2017, and 2022, respectively. He was awarded a Ph.D. fellowship by the Spanish Ministry of Education in September 2018. He is currently an Interim Assistant Professor with the Department of Signal Theory, Telematics, and Communications, University of Granada (Spain), and has been a Visiting Researcher at NEC Laboratories Europe (Germany) on several occasions. His research interests include network dimensioning in 5G/6G RAN, 5G/6G integration with TSN, and quantum communications and their integration with classical networks, with a focus on mathematical modeling and architecture design across these domains.
\end{IEEEbiography}
\vskip -2\baselineskip plus -1fil 
\begin{IEEEbiography}
[{\includegraphics[width=1in,height=1.25in,clip,keepaspectratio]{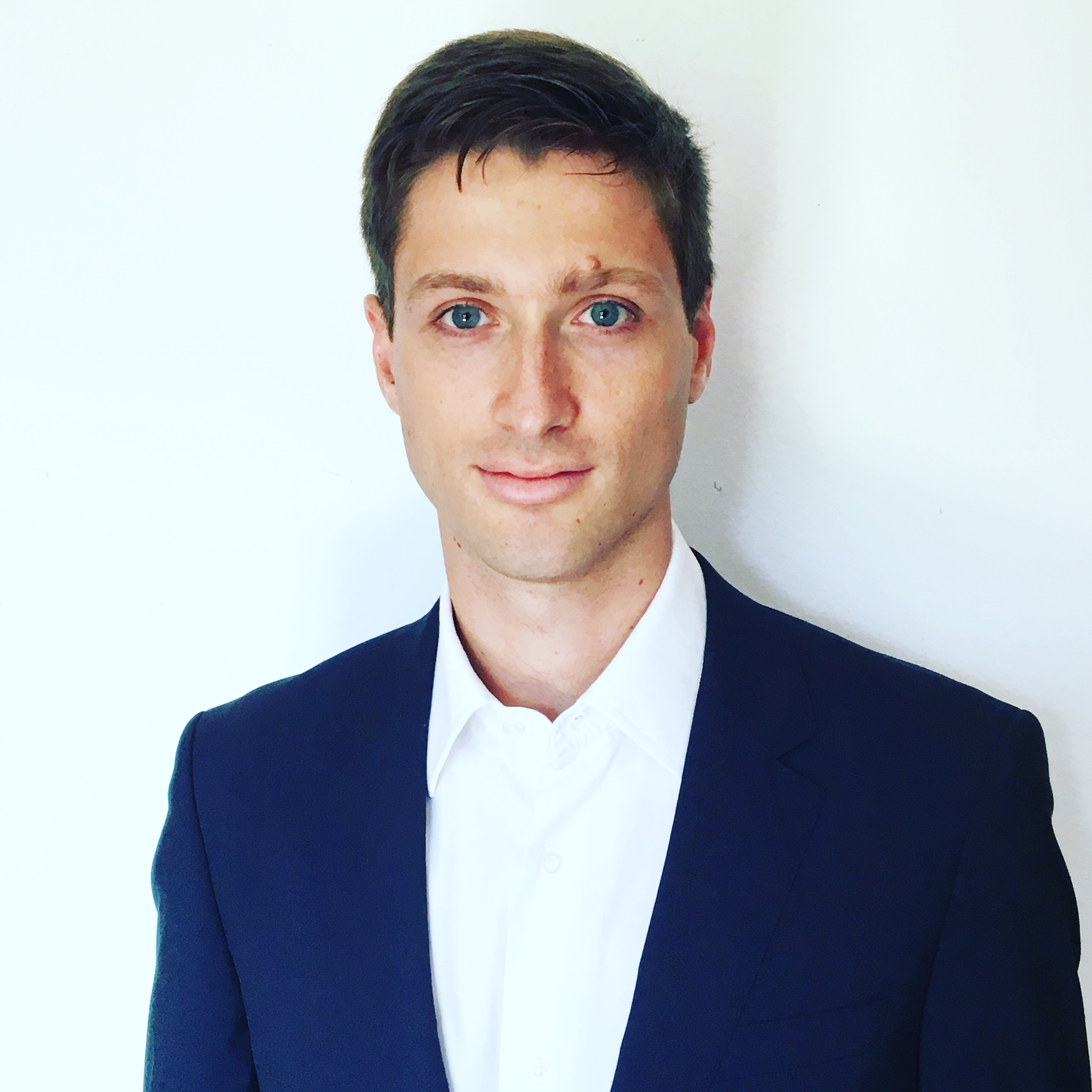}}]
{Lanfranco Zanzi } (Member, IEEE) received the B.Sc. and M.Sc. degrees in telecommunication engineering from the Polytechnic of Milan, Italy, in 2014 and 2017, respectively, and the Ph.D. degree from the Technical University of Kaiserlautern, Germany, in 2022. He works as a Senior Research Scientist with NEC Laboratories Europe. His research interests include network virtualization, machine learning, blockchain, and their applicability to 5G and 6G mobile networks.
\end{IEEEbiography}
\vskip -2\baselineskip plus -1fil
\begin{IEEEbiography}
[{\includegraphics[width=1in,height=1.25in,clip,keepaspectratio]{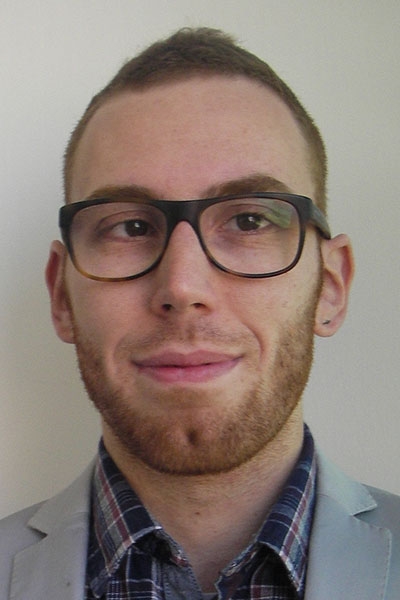}}]
{Vincenzo Sciancalepore } (Senior Member, IEEE) is a Principal Researcher with NEC
Laboratories Europe GmbH, Germany. He is currently focusing his activity in the area of reconfigurable intelligent surfaces (RIS) and integrated sensing and communications. He is the Standard Delegate of NEC actively contributing
to the standard ETSI RIS ISG. He currently holds the Italian habilitation as a Full Professor of Telecommunications issued by MIUR. He is currently the Innovation Project Manager of an H2020-fundeed project called 6G-GOALS that will analyze and design new innovation for semantic communications on 6G systems. He is an Editor of IEEE TRANSACTIONS ON WIRELESS COMMUNICATIONS and
IEEE TRANSACTIONS ON COMMUNICATIONS. He is currently a member of the IEEE Emerging Technologies Standing Committee leading the initiatives on RISs.
\end{IEEEbiography}
\vskip -2\baselineskip plus -1fil
\begin{IEEEbiography}
[{\includegraphics[width=1in,height=1.25in,clip,keepaspectratio]{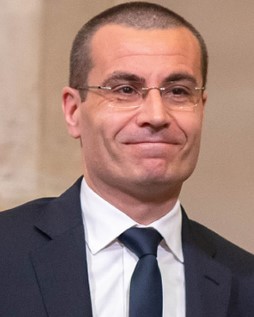}}]
{Marco Di Renzo } (Fellow, IEEE) received the Laurea (cum laude) and Ph.D. degrees in electrical engineering from the University of L’Aquila, Italy, in 2003 and 2007, respectively, and the Habilitation à Diriger des Recherches (Doctor of Science) degree from University Paris-Sud (currently Paris-Saclay University), France, in 2013. Currently, he is Chair Professor of Telecommunications Engineering, the Director of the Centre for Telecommunications Research, and the Head of the Telecommunications Group, Department of Engineering, King’s College London, London, United Kingdom. He is also a CNRS Research Director (Professor) with the Laboratory of Signals and Systems at CNRS-CentraleSupélec, Paris-Saclay University, Paris, France. He was a France-Nokia Chair of Excellence in ICT at the University of Oulu (Finland), a Tan Chin Tuan Exchange Fellow in Engineering at Nanyang Technological University (Singapore), a Fulbright Fellow at The City University of New York (USA), a Nokia Foundation Visiting Professor at Aalto University (Finland), and a Royal Academy of Engineering Distinguished Visiting Fellow at Queen’s University Belfast (U.K.). He is a Fellow of the IEEE, IET, EURASIP, and AAIA; an Academician of AIIA; an Ordinary Member of the European Academy of Sciences and Arts, an Ordinary Member of the Academia Europaea, and an Ordinary Member of the Italian Academy of Technology and Engineering; an Ambassador of the European Association on Antennas and Propagation; and a Highly Cited Researcher. He has received several distinctions, including the Michel Monpetit Prize conferred by the French Academy of Sciences, the IEEE Communications Society Heinrich Hertz Award, and the IEEE Communications Society Marconi Prize Paper Award in Wireless Communications. Also, he is a principal investigator of an ERC Synergy grant on metasurface-based information processing. He served as the Editor-in-Chief of IEEE Communications Letters from 2019 to 2023, and as the Director of Journals and Chair of the Publications Misconduct Ad Hoc Committee of the IEEE Communications Society from 2024 to 2025. Currently, he sits on the IEEE-COMSOC Fellow Evaluation Standing Committee and on the Editorial Board of the Proceedings of the IEEE.
\end{IEEEbiography}
\vskip -2\baselineskip plus -1fil
\begin{IEEEbiography}
[{\includegraphics[width=1in,height=1.25in,clip,keepaspectratio]{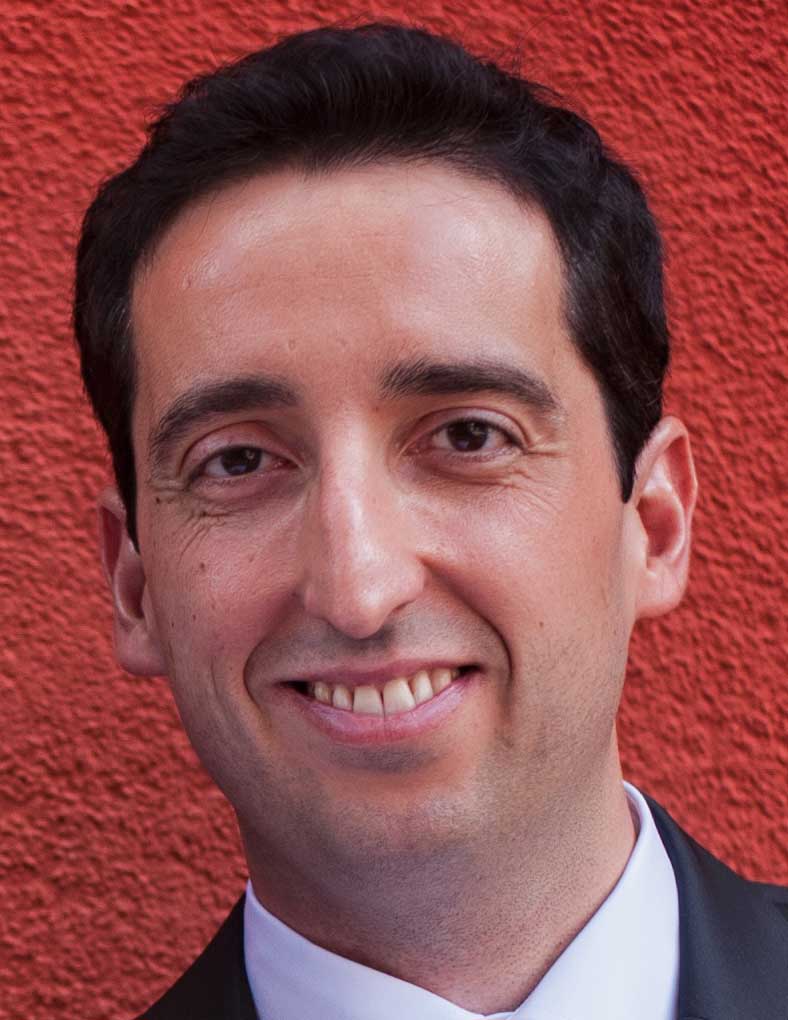}}]
{Xavier~Costa-P\'erez } (Senior Member, IEEE) received the Ph.D. degree in telecommunications from the Polytechnic University of Catalonia, Barcelona. He is a Research Professor with ICREA, a Scientific Director with i2Cat Research Center, and the Head of 5G/6G Networks R\&D with NEC Laboratories Europe. He has served on the Organizing Committees of several conferences, published papers of high impact, and holds tenths of granted patents. He was a recipient of the National Award for his Ph.D. thesis.
\end{IEEEbiography}
\vskip -2\baselineskip plus -1fil

\end{document}